\documentclass[onecolumn,titlepage,showpacs]{revtex4}

\begin{document}


\title{RECENT RESEARCH DEVELOPMENTS IN PHYSICS\bigskip

Gauge theories of gravity: the nonlinear framework}


\author{A. Tiemblo}
\email[]{laetr03@imaff.cfmac.csic.es}

\author{R. Tresguerres}
\email[]{romualdo@imaff.cfmac.csic.es}
\affiliation{Instituto de Matem\'aticas y F\'isica Fundamental\\
Consejo Superior de Investigaciones Cient\'ificas\\ Serrano 113
bis, 28006 Madrid, SPAIN}

\date{\today}

\begin{abstract}
Nonlinear realizations of spacetime groups are presented as a
versatile mathematical tool providing a common foundation for
quite different formulations of gauge theories of gravity. We
apply nonlinear realizations in particular to both the Poincar\'e
and the affine group in order to develop Poincar\'e gauge theory
(PGT) and metric-affine gravity (MAG) respectively. Regarding PGT,
two alternative nonlinear treatments of the Poincar\'e group are
developed, one of them being suitable to deal with the Lagrangian
and the other one with the Hamiltonian version of the same gauge
theory. We argue that our Hamiltonian approach to PGT is closely
related to Ashtekar's approach to gravity. On the other hand, a
brief survey on MAG clarifies the role played by the
metric--affine metric tensor as a Goldsone field. All
gravitational quantities in fact --the metric as much as the
coframes and connections-- are shown to acquire a simple
gauge--theoretical interpretation in the nonlinear framework.
\end{abstract}

\pacs{04.50.+h, 11.15.-q}
\maketitle

\section{Introduction}

In the search for the unification of forces, different
alternatives to Einstein's original General Relativity have been
proposed \cite{Utiyama:1956sy} \cite{Sciama1} \cite{Sciama2}
\cite{Kibble:1961ba} \cite{Hayashi:1980wj} \cite{Ivanenko:1983vf}
\cite{Lord:1987uq} \cite{Lord:1988nd} \cite{Sardanashvily:2002mi}
\cite{Ashtekar:1986yd} \cite{Ashtekar:1987gu}, based on the
extension of the gauge principle to spacetime groups. The analogy
existing between gauge theories of gravity and the Yang--Mills
theories supporting the standard model allows to assimilate
gravitation to the remaining forces through the characterization
of all interactions --gravity included-- as mediated by gauge
potentials only. In what follows we will focus our attention on
Hehl's Poincar\'e gauge theory (PGT) as much as on metric-affine
gravity (MAG), see \cite{Hehl:1974cn} \cite{Hehl:1976kj}
\cite{Hehl:1990yq} \cite{Gronwald:1995em} \cite{Hehl:1995ue}, both
formalisms presenting grand adaptability in dealing with a
diversity of spacetime actions. We claim their value as a suitable
support for the unification of different theoretical points of
view on gravitational forces. Indeed, a main result of the present
work is to show the close relationship between the Hamiltonian
version of PGT and the Ashtekar approach.

The cornerstone of our treatment consists of nonlinear
realizations (NLR's), a mathematical method \cite{Coleman:1969sm}
\cite{Callan:1969sn} \cite{Salam:1969rq} \cite{Isham:1971dv}
\cite{Borisov74} \cite{Cho:1978ss} \cite{Stelle:1980aj} argued by
us to provide a universal foundation for gauge theories of
different groups \cite{Julve:1994bh} \cite{Julve:1995yf}
\cite{Lopez-Pinto:1995qb} \cite{Lopez-Pinto:1996pu}
\cite{Lopez-Pinto:1997aw} \cite{Tresguerres:2000qn}
\cite{Tiemblo03}. The usefulness of NLR's in the context of
gravitational theories becomes apparent mainly when translations
are contained in the spacetime gauge group as a subgroup. As a
result of the nonlinear approach, the translational connections
transform into covector-valued 1--forms suitable to be identified
as coframes, so that the dynamical gauge theory becomes
indistinguishable from spacetime geometry.

For a gauge group $G$ to be realized nonlinearly, an auxiliary
subgroup $H\subset G$ is required to be chosen in addition. The
freedom in selecting the latter provides the nonlinear method with
a considerable flexibility. Indeed, a single theory, say the gauge
theory of the Poincar\'e group $G$, manifests itself in quite
different forms depending on the subgroup $H$ chosen. As we will
see, the NLR of PGT with $H={\rm Lorentz}$ reveals to be suitable
to be taken as the basis for a Lagrangian approach, whereas the
one with $H=SO(3)$ is especially adapted to a Hamiltonian
treatment. In the case of MAG, being $G$ the affine group, we also
consider two different nonlinear approaches, corresponding to the
choices of the subgroups $H=GL(4\,,R)$ and $H={\rm Lorentz}$
respectively. The relation between both NLR's will be exploited to
explain the gauge theoretical origin of the ten degrees of freedom
of the metric, as much as their Goldsone nature allowing to
rearrange them into redefined fields.

The present work is organized as follows. In Section {\bf II} we
briefly review the mathematical foundations of NLR's in terms of
composite fiber bundles. Section {\bf III} is devoted to what we
call the standard approach to PGT, yielding explicitly Lorentz
covariant coframes and spin connections. The resulting formalism
is used to build an Einstein--Cartan Lagrangian description of
gravity. Then in Section {\bf IV} we pay attention to the
Hamiltonian approach to PGT. Several technical aspects are
discussed, such as a Poincar\'e invariant foliation of spacetime
and a general Hamiltonian formalism adapted to exterior calculus.
The Hamiltonian dynamics of the Einstein--Cartan action, expressed
in terms of real PGT connection variables, is shown to be
consistent both with the Lagrangian treatment of Section {\bf III}
and with an alternative approach of the Ashtekar type which we
also develop in some detail. Finally in Section {\bf V} we apply
the nonlinear approach to the affine group to derive
metric--affine gravity.

\section{Foundations of nonlinear gauge theories}

Principal bundles $P(M\,,G)$ describe the structure of ordinary
gauge theories of internal Lie groups $G$. This scheme does not
hold for nonlinear gauge theories, based on the interplay between
the gauge group $G$ and a subgroup $H\subset G$. In
\cite{Tresguerres:2002uh} we invoked composite fiber bundles as
the suitable topological background underlying nonlinear
realizations of local symmetries.

Roughly speaking, a composite bundle is a principal bundle
$P(M\,,G)$ whose $G$-diffeomorphic fibers are regarded themselves
as bundles whose structure group $H$ is a subgroup of the
structure group $G$ of the total bundle. Actually, composite
bundles can be built provided a subgroup $H\subset G$ exists,
whose right action on elements $g\in G$ induces a complete
partition of the group manifold $G$ into mutually disjoint orbits
$gH$. By projecting each of these equivalence classes to a single
element (that is to a left coset) of the quotient space $G/H$, the
group manifold $G$ becomes organized as a bundle $G(G/H\,,H)$ with
the orbits $gH$ (diffeomorphic to the subgroup $H$) as local
fibers and with $G/H$ as the base space. When attached to points
of an auxiliary base manifold $M$, the local bundles $G(G/H\,,H)$
constitute the fibers of a composite bundle.

Locally, composite bundles are diffeomorphic to $M\times G/H\times
H$, so that by singling out either the base space $M$ or the
manifold $\Sigma\simeq M\times G/H$, two mutually related bundle
structures can be recognized in the total bundle space $P$. On the
one hand, the usual bundle structure survives with total
$G$-diffeomorphic fibers projecting to the bundle base space $M$.
On the other hand, $P$ can also be regarded as consisting of
$H$-diffeomorphic fiber branches attached to the manifold
$\Sigma$, the latter playing the role of an {\it intermediate base
space}. The alternative projections $\pi _{_{PM}}:P\rightarrow M$
and $\pi _{_{P\Sigma}}:P\rightarrow\Sigma\,$ become related to
each other by defining an additional mapping $\pi _{_{\Sigma
M}}:\Sigma\rightarrow M\,$ such that the ordinary total projection
decomposes into two partial projections as $\pi _{_{PM}}=\pi
_{_{\Sigma M}}\circ\pi _{_{P\Sigma}}\,$. Correspondingly, the
local sections $s_{_{MP}}:M\rightarrow P$ decompose as
\begin{equation}
s _{_{MP}}=s _{_{\Sigma P}}\circ s
_{_{M\Sigma}}\,.\label{sectdecomp}
\end{equation}

Let us express the sections introduced in (\ref{sectdecomp}) in
terms of zero sections --the one associated to $s_{_{MP}}$ denoted
as $\sigma _{_{MP}}$ and so on--, that is
\begin{equation}
s_{_{MP}} =R_{\tilde{g}}\circ\sigma _{_{MP}}\,,\quad \tilde{g}\in
G \,,\label{locsect1}
\end{equation}
\begin{equation}
s_{_{\Sigma P}} =R_a\circ\sigma _{_{\Sigma P}}\,,\quad a\in H
\,,\label{locsect2}
\end{equation}
and
\begin{equation}
s_{_{M\Sigma}} =R_b\circ\sigma _{_{M\Sigma}}\,,\quad b\in G/H
\,.\label{locsect3}
\end{equation}
As compatibility conditions, they have to satisfy $\tilde{g}
=b\cdot a$ and $\sigma _{_{MP}}=R_{b^{-1}}\circ\sigma _{_{\Sigma
P}}\circ R_b\circ\sigma _{_{M\Sigma}}\,$ (or alternatively $\sigma
_{_{\Sigma P}}\circ R_b =R_b\circ\sigma _{_{\Sigma P}}\,$). For
later convenience we introduce the composite section $\sigma
_{\xi}:M\rightarrow\Sigma\rightarrow P$ defined from the total and
zero sections in (\ref{locsect2}) and (\ref{locsect3}) as
\begin{equation}
\sigma _{\xi}(x):=\sigma _{_{\Sigma P}}\circ
s_{_{M\Sigma}}(x)=R_b\circ\sigma _{_{MP}}(x)\,.\label{sigmaxi}
\end{equation}
(The $\xi$ in $\sigma _{\xi}(x)$ stands for the parameters
labelling the elements $b\in G/H$ displayed as $R_b$ in the r.h.s.
of (\ref{sigmaxi}).) The reason for introducing (\ref{sigmaxi}) is
its usefulness for expressing the main results on the nonlinear
approach, deduced in \cite{Tresguerres:2002uh} and summarized
below.

By comparing two bundle elements, both of the form
(\ref{locsect2}), differing from each other by the left action
$L_g$ of elements $g\in G$, in \cite{Tresguerres:2002uh} we found
the nonlinear transformation law
\begin{equation}
L_g\circ\sigma _\xi (x)=R_h\circ\sigma _{\xi
'}(x)\,,\label{nonlintrans2}
\end{equation}
being $R_h$ the right action of a certain element $h\in H$ of the
subgroup. For practical reasons, in \cite{Tiemblo03} we
transformed (\ref{nonlintrans2}) into the more manageable formula
\begin{equation}
g\cdot b =b\,'\cdot h\,,\label{simplif}
\end{equation}
where $g\in G$, $h\in H$, and $b,b\,'\in G/H$. Notice that
(\ref{simplif}) reproduces the original form of the nonlinear law
as given in \cite{Coleman:1969sm}. On the other hand, for
infinitesimal $g$ and $h=e^\mu\approx 1+\mu$, a gauge
transformation is induced by (\ref{nonlintrans2}) on fields $\psi$
of any representation space of $H$, namely
\begin{equation}
\delta\psi (\sigma _\xi (x))=\rho (\mu)\psi (\sigma _\xi
(x))\,,\label{varmatt}
\end{equation}
where $\rho (\mu)$ denotes the suitable representation of the
$H$-algebra, see \cite{Coleman:1969sm} \cite{Tresguerres:2002uh}.

Since we are interested in building the covariant derivatives of
the fields $\psi$ transforming nonlinearly as (\ref{varmatt}), in
\cite{Tiemblo03} we compared the ordinary linear connection,
resulting from pulling back the connection 1-form $\omega$ by
means of $\sigma _{_{MP}}$, that is
\begin{equation}
A_{_M} =\sigma _{_{MP}}^*\,\omega\,,\label{linconn1}
\end{equation}
with the connection characteristic for the nonlinear approach,
defined as the pullback of $\omega$ by means of $\sigma _\xi$,
namely
\begin{equation}
\Gamma _{_M}=\sigma _\xi ^*\,\omega\,,\label{nlinconn3}
\end{equation}
the difference between $\sigma _\xi$ and $\sigma _{_{MP}}$ being
displayed in (\ref{sigmaxi}). One finds \cite{Tiemblo03} that the
nonlinear connection (\ref{nlinconn3}) can be expressed in terms
of the linear one (\ref{linconn1}) as
\begin{equation}
\Gamma _{_M}=b^{-1}(\,d + A_{_M} )\,b\,.\label{compar4}
\end{equation}
Its gauge transformations, induced by the nonlinear group action
(\ref{nonlintrans2}), are found to be
\begin{equation}
\delta\Gamma _{_M}= -(\,d\mu +[\Gamma _{_M}\,,\mu
]\,)\,,\label{varnlcon}
\end{equation}
with $\mu$ the same $H$-algebra-valued parameters as in
(\ref{varmatt}). Being $\Gamma _{_M}$ valued on the Lie algebra of
the whole group $G$, from (\ref{varnlcon}) one reads out that only
those of its components defined on the $H$ algebra behave as true
connections transforming inhomogeneously, while its components
with values on the remaining algebra elements of $G/H$ transform
as $H$-tensors. As a result of (\ref{varmatt}) and
(\ref{varnlcon}), covariant differentials defined as
\begin{equation}
D\psi :=(d +\rho (\Gamma _{_M}))\psi\,,\label{gencovder}
\end{equation}
transform as
\begin{equation}
\delta D\psi =\rho (\mu )D\psi\,.\label{vargencovder}
\end{equation}

Finally, let us make use of the covariant differential operator
\begin{equation}
D:=d +\Gamma _{_M}\,,\label{covdiff}
\end{equation}
as read out from (\ref{gencovder}) without specifying any
particular representation, to obtain the field strength as
\begin{equation}
F:=D\wedge D=\,d\,\Gamma _{_M} +\Gamma _{_M}\wedge \Gamma
_{_M}\,.\label{fieldstrength}
\end{equation}
In view of (\ref{varnlcon}) we find (\ref{fieldstrength}) to
transform as
\begin{equation}
\delta F =[\,\mu\,,F ]\,.\label{fvariat}
\end{equation}

The relevance of the nonlinear approach for the foundation of
gauge theories of gravity becomes evident in the following
sections, where we apply it to the local treatment of different
spacetime groups.

\section{\bf Standard nonlinear Poincar\'e gauge theory}

\subsection{\bf Coframes and Lorentz connections}

As a first application of the general formalism established in
previous section, let us take the group $G$ to be the Poincar\'e
group in order to show how its nonlinear local approach gives rise
to the Poincar\'e gauge theory of gravity (PGT). Diverse nonlinear
realizations are possible depending on the choice of the auxiliary
subgroup $H\subset G$. In this section we take $H$ to be the
Lorentz group, yielding an explicitly Lorentz covariant
four--dimensional formalism which provides the geometrical basis
for a Lagrangian approach (developed by us in the language of
exterior calculus). Later in Section IV we present the version of
PGT resulting from taking $H$ to be $SO(3)$, suitable to deal with
the $3+1$ decomposition underlying PGT Hamiltonian dynamics (to be
treated also in exterior calculus, with differential forms playing
the role of dynamical variables).

Our starting point is the fundamental transformation law
(\ref{nonlintrans2}) of nonlinear realizations. After rewriting it
in the simplified form (\ref{simplif}), for $G=$ Poincar\'e and
$H=$ Lorentz we parametrize the infinitesimal Poincar\'e group
element $g\in G$ and the infinitesimal Lorentz group element $h\in
H$ respectively as
\begin{equation}
g =\,e^{i\,\epsilon^\alpha P_\alpha } e^{i\,\beta
^{\alpha\beta}L_{\alpha\beta}} \approx\,1+i\,\left(
\epsilon^\alpha P_\alpha +\beta
^{\alpha\beta}L_{\alpha\beta}\,\right)\,,\label{ginfin}
\end{equation}
and
\begin{equation}
h =\,e^{i\,\mu ^{\alpha\beta}L_{\alpha\beta}} \approx\,1+i\,\mu
^{\alpha\beta}L_{\alpha\beta}\,,\label{hinfin}
\end{equation}
where $L_{\alpha\beta}$ are the Lorentz generators and $P_\mu$ the
translational ones. As read out from (\ref{simplif}), the left
action of (\ref{ginfin}) on elements
\begin{equation}
b=e^{-i\,\xi ^\alpha P_\alpha}\label{bfinite}
\end{equation}
of the coset space $G/H$, being (\ref{bfinite}) identical with
elements of the group of translations labelled by the finite
parameters $\xi ^\alpha$, induces a right action of (\ref{hinfin})
on
\begin{equation}
b\,'\,=e^{-i\,(\xi ^\alpha +\delta\xi ^\alpha )
P_\alpha}\,.\label{bfinitetransf}
\end{equation}
Replacing (\ref{ginfin})--(\ref{bfinitetransf}) into
(\ref{simplif}) and taking into account the commutation relations
of the Poincar\'e algebra
\begin{eqnarray}
\left[P_{\alpha}\, , P_{\beta}\right] &=& 0\,,\nonumber \\
\left[L_{\alpha\beta }\,, P_\mu\right] &=& i\,o_{\mu [\alpha
}P_{\beta ]}\,,\nonumber \\
\left[L_{\alpha\beta }\,,L_{\mu\nu }\right] &=& -i\,\left(
o_{\alpha [\mu } L_{\nu ]\beta} - o_{\beta [\mu }L_{\nu ]\alpha
}\right)\,,\label{comrelpoinc}
\end{eqnarray}
with $o_{\alpha\beta}$ as the the Minkowski metric
\begin{equation}
o_{\alpha\beta}:=diag\,(-+++)\,,\label{metric}
\end{equation}
a simple computation with the help of the Hausdorff-Campbell
formula, see Appendix {\bf B}, yields the value of $\mu
^{\alpha\beta}$ in (\ref{hinfin}) as much as the variation of the
translational coset parameters, namely
\begin{equation}
\mu ^{\alpha\beta}=\beta ^{\alpha\beta}\,,\qquad \delta \xi
^\alpha =-\beta _\beta {}^\alpha\,\xi ^\beta -\epsilon
^\alpha\,,\label{paramvars}
\end{equation}
showing $\xi ^\alpha$ to transform exactly as Minkowskian
coordinates.

On the other hand, using for the linear connection
(\ref{linconn1}) of the Poincar\'e group the notation
\begin{equation}
A_{_M}=-i\,{\buildrel (T)\over{\Gamma ^\alpha}} P_\alpha
-i\,\Gamma ^{\alpha\beta}L_{\alpha\beta}\,,\label{linconnpoinc}
\end{equation}
whose components on the Poincar\'e algebra are the linear
translational contribution ${\buildrel (T)\over{\Gamma ^\alpha}}$
and the Lorentz one $\Gamma ^{\alpha\beta}$ respectively, we find
the nonlinear connection (\ref{compar4}) to be
\begin{equation}
\Gamma _{_M}=-i\,\vartheta ^\alpha P_\alpha -i\,\Gamma
^{\alpha\beta}L_{\alpha\beta}\,,\label{nonlinconnpoinc}
\end{equation}
with the Lorentz connection $\Gamma ^{\alpha\beta}$ unmodified
with respect to the linear case (\ref{linconnpoinc}), but with the
translational connection transformed into
\begin{equation}
\vartheta ^\alpha :=\,D\,\xi ^\alpha +{\buildrel (T)\over{\Gamma
^\alpha}}\,.\label{tetrad}
\end{equation}
In view of (\ref{varnlcon}), the components of
(\ref{nonlinconnpoinc}) transform respectively as
\begin{equation}
\delta\vartheta ^\alpha =-\vartheta ^\beta\beta _\beta
{}^\alpha\,,\qquad \delta\Gamma ^{\alpha\beta} =D\beta
^{\alpha\beta}\,.\label{varsconns}
\end{equation}
The most relevant result is the first equation in
(\ref{varsconns}). According to it, instead of the linear
translational connection ${\buildrel (T)\over{\Gamma ^\alpha}}$ in
(\ref{linconnpoinc}) transforming inhomogeneously as $\delta
{\buildrel (T)\over{\Gamma ^\alpha}}=-{\buildrel (T)\over{\Gamma
^\beta}}\beta _\beta{}^\alpha +D\epsilon ^\alpha$, we have at our
disposal a nonlinear translational connection 1-form
(\ref{tetrad}) which is Lorentz covector-valued. The latter will
be identified from now on as the Lorentz coframe or tetrad. This
feature of deductively providing tetrads with the right
transformation properties constitutes one of the main achievements
of nonlinear realizations. (Compare with the hypotheses needed to
build (\ref{tetrad}) in the context of the linear approach
\cite{Mielke:1993gk}.)

\subsection{\bf Gravitational actions and Lagrangian field equations
(in the language of exterior calculus)}

Let us show that the coframe $\vartheta ^\alpha$ and the Lorentz
connection $\Gamma ^{\alpha\beta}$ introduced above are variables
suitable to build Poincar\'e gauge invariant gravitational
actions. (Exterior calculus allows to take differential forms as
such --rather than their components-- as dynamical variables.
Actually we obtain the field equations by varying a Lagrangian
density 4--form with respect to the 1--forms $\vartheta ^\alpha$
and $\Gamma ^{\alpha\beta}$ respectively, see below and
\cite{Hehl:1995ue}.) The field strengths of the coframe and of the
Lorentz connection are found by applying the general expression
(\ref{fieldstrength}) to the nonlinear Poincar\'e connection
(\ref{nonlinconnpoinc}), yielding
\begin{equation}
F=-i\,T^\alpha P_\alpha
-i\,R^{\alpha\beta}L_{\alpha\beta}\,.\label{poincfielstr}
\end{equation}
In (\ref{poincfielstr}), the torsion
\begin{equation}
T^\alpha :=D\vartheta ^\alpha :=d\vartheta ^\alpha +\Gamma
_\beta{}^\alpha\wedge\vartheta ^\beta\label{torsion}
\end{equation}
coincides with the translational field strength, while the
Lorentzian field strength is the curvature
\begin{equation}
R_\alpha{}^\beta :=d\Gamma _\alpha{}^\beta +\Gamma
_\gamma{}^\beta\wedge\Gamma _\alpha{}^\gamma\,.\label{curvature}
\end{equation}
Both (\ref{torsion}) and (\ref{curvature}) are building blocks for
gravitational actions. For instance, with the help of
(\ref{curvature}) besides the elements of the eta basis defined in
Appendix {\bf A} (built from the coframes (\ref{tetrad})), one can
express the ordinary Einstein-Cartan gravitational Lagrange
density 4--form with cosmological term as
\begin{equation}
L_{EC}=-{1\over{2l^2}}\,R^{\alpha\beta}\wedge\eta
_{\alpha\beta}+{\Lambda\over{l^2}}\,\eta\,,\label{EinstCartlagr}
\end{equation}
see \cite{Hehl:1995ue} \cite{Obukhov:1996eq}. More general
Lagrangians including contributions quadratic in the irreducible
pieces of curvature and torsion (of the form
$^{(I)}R^{\alpha\beta}\wedge\,^*R_{\alpha\beta}$,
$^{(I)}T^\alpha\wedge\,^*T_\alpha$) are extensively studied in the
literature \cite{Hecht:1995fq} \cite{Obukhov:1996eq}. For the sake
of simplicity, here we only consider the action $S=\int L_{EC}\,$
built from (\ref{EinstCartlagr}). The field equations derived by
varying (\ref{EinstCartlagr}) with respect to the tetrads
$\vartheta ^\alpha$ are
\begin{equation}
{1\over 2}\,\eta _{\alpha\beta\gamma}\wedge
R^{\beta\gamma}-\Lambda\,\eta _\alpha =0\,,\label{Einsteqn}
\end{equation}
and on the other hand, variation on the Lorentz connection $\Gamma
^{\alpha\beta}$ yields
\begin{equation}
D\eta _{\alpha\beta}=0\,.\label{zerotorsion}
\end{equation}
Since $D\eta _{\alpha\beta}=\eta _{\alpha\beta\gamma}\wedge
T^\gamma$, from (\ref{zerotorsion}) follows the vanishing of
torsion, that is
\begin{equation}
T^\alpha =\,0\,,\label{zerotorsionbis}
\end{equation}
implying that the Lorentz connection reduces to the (anholonomic)
Christoffel connection
\begin{equation}
\Gamma ^{\{\}}_{\alpha\beta}:=\, e_{[\alpha }\rfloor d\,\vartheta
_{\beta ]} -{1\over2} \left( e_\alpha\rfloor e_\beta\rfloor
d\,\vartheta ^\gamma\right) \vartheta _\gamma
\,.\label{Christoffel}
\end{equation}

By replacing (\ref{Christoffel}) in (\ref{Einsteqn}), the latter
reduces to the standard Einstein vacuum equations with
cosmological constant defined on a Riemannian space. This can be
easily checked by translating (\ref{Einsteqn}) to the usual
Riemannian language of General Relativity involving the holonomic
metric $g_{\,ij}:=o_{\alpha\beta}\,e_i{}^\alpha e_j{}^\beta\,$
defined from the tetrads $\vartheta ^\alpha =dx^i e_i{}^\alpha$
with the Minkowski metric (\ref{metric}). The anholonomic
Christoffel connection (\ref{Christoffel}) transforms into
\begin{equation}
\Gamma ^{\{\}}_{\alpha\beta}:=-\,dx^i\, e_{[\alpha
}{}^j\,\Bigl(\,\partial _i e_{j\beta ]}-\Gamma _{ij}{}^k e_{k\beta
]}\,\Bigr)\label{standardChristoffel}
\end{equation}
when reexpressed in terms of the ordinary holonomic Christoffel
symbol $\Gamma _{ij}{}^k :={1\over 2}\,g^{\,kl}\Bigl(\,\partial _i
g_{\,lj}+
\partial _j g_{\,li}-\partial _l g_{\,ij}\,\Bigr)$, while the curvature
(\ref{curvature}) with (\ref{standardChristoffel}) reduces to
\begin{equation}
R_{\alpha\beta}={1\over 2}\,dx^i\wedge dx^l\,e_{[\alpha
}{}^j\,e_{k\beta ]}\,R_{\,ilj}{}^k\,,\label{Riemann}
\end{equation}
being $R_{\,ilj}{}^k :=2\,\Bigl(\,\partial _{[i}\Gamma _{l]j}{}^k
+\Gamma _{[im}{}^k\Gamma _{l]j}{}^m\,\Bigr)$ the ordinary Riemann
tensor. By inserting (\ref{Riemann}) into (\ref{Einsteqn}), using
the definitions of the Ricci tensor $R_{\,ij}:=R_{\,ikj}{}^k$ and
of the scalar curvature $R:=g^{\,ij} R_{\,ij}\,$ and making use of
the holonomic version of the eta basis of Appendix {\bf A}, being
for instance $\eta ^j =\eta ^\alpha\,e_{\alpha}{}^j
={1\over{3!}}\,\sqrt{g}\,\,\epsilon ^j{}_{klm}\,dx^k\wedge
dx^l\wedge dx^m\,$, the Einstein equations (\ref{Einsteqn}) take
their standard form
\begin{equation}
{1\over 2}\,\eta _{\alpha\beta\gamma}\wedge
R^{\beta\gamma}-\Lambda\,\eta _\alpha =-\,e^i{}_\alpha\,\Bigl(\,
R_{\,ij}-{1\over 2}\, g_{\,ij}\,R +\Lambda\,g_{\,ij}\,\Bigr)\eta
^j\,=0\,.\label{standardEinsteqn}
\end{equation}
The fact that the Einstein-Cartan action (\ref{EinstCartlagr})
reproduces the Einstein equations of General Relativity is a test
of the validity of the PGT approach. However, the latter is
flexible enough to be applied to extended gravitational actions
involving quadratic curvature and torsion terms and giving rise to
nonvanishing torsion. In general, the use of Poincar\'e gauge
variables introduces both, a different perspective in the
interpretation of gravity as mediated by connections only
--translational as much as Lorentzian ones-- rather than by the
base space metric $g_{\,ij}$, and moreover the possibility of
deriving new results which are meaningless in the purely metrical
approach. In order to illustrate the latter point, in the next
paragraph we show as an example the coupling of gravitational
fields (namely $\vartheta ^\alpha$ and $\Gamma ^{\alpha\beta}$) to
fundamental matter fields.

\subsection{PGT invariant action of Dirac fields}

If PGT is to be regarded as a basic theory of gravity, one has to
understand its coupling to matter beyond phenomenological matter
sources. Accordingly, a PGT invariant Dirac action is to be added
to PGT gravitational Lagrangians like (\ref{EinstCartlagr}) or its
generalizations. To do so, the first step consists in finding the
explicit form of the covariant derivative (\ref{gencovder}) of
Dirac bispinors with the Poincar\'e nonlinear connection
(\ref{nonlinconnpoinc}). As shown in \cite{Tiemblo03}, a
four--dimensional realization of the Poincar\'e generators,
$P_\mu$ as much as $L_{\alpha\beta}$, can be built from the gamma
matrices in the Dirac representation
\begin{equation}
\gamma ^0 =\left(\begin{array}{ll}
I&\,\,0\\
0&-I
\end{array}\right)\,,\,
\gamma ^a =\left(\begin{array}{ll}
\,\,0&\sigma ^a\\
-\sigma ^a&0
\end{array}\right)\,,\,
\gamma _5 =\left(\begin{array}{ll}
0&I\\
I&0
\end{array}\right)\,,\label{matrices}
\end{equation}
being $\gamma _5:=i\,\gamma ^0\gamma ^1\gamma ^2\gamma ^3$.
Besides the usual spinor generators
\begin{equation}
\rho ( L_{\alpha\beta})=\sigma
_{\alpha\beta}:={i\over8}\,[\,\gamma _\alpha\,,\gamma _\beta
]\,,\label{spinorgens}
\end{equation}
we introduce the finite matrix representation of translational
generators as
\begin{equation}
\rho (P_\mu ) =\pi _\mu :={m\over4}\,\gamma _\mu (\,1+\gamma _5
)\,,\label{linmom2}
\end{equation}
where the dimensional constant $m\sim [L]^{-1}$ (in natural units
$\hbar =c=1$) guarantees the same dimensionality for the intrinsic
linear momentum associated to (\ref{linmom2}) as for the orbital
linear momentum $-i\partial _\mu$, see \cite{Tiemblo03}. Both
(\ref{spinorgens}) and (\ref{linmom2}) provide a nontrivial finite
matrix realization of the Poincar\'e algebra (\ref{comrelpoinc})
in spite of the fact that $\pi _\mu \pi _\nu =0$. The Poincar\'e
covariant derivative (\ref{gencovder}) of Dirac fields thus reads
\begin{equation}
D\psi =d\psi -i\,(\Gamma ^{\alpha\beta}\sigma _{\alpha\beta}
+\vartheta ^\mu \pi _\mu )\psi\,,\label{covder1}
\end{equation}
transforming in accordance with (\ref{vargencovder}) as $\delta
D\psi =i\,\beta ^{\alpha\beta}\sigma _{\alpha\beta}D\psi$. The
PGT--invariant Dirac Lagrange density 4-form built with the help
of (\ref{covder1}) --without explicit mass term-- reads
\begin{equation}
L_D={i\over2}\,(\overline{\psi}\,\,^*\gamma\wedge D\psi
+\overline{D\psi}\wedge\,^*\gamma\psi )\,,\label{mattlagrang}
\end{equation}
where we use the notation of \cite{Hehl:1990yq}, being $\gamma
:=\vartheta ^\mu \gamma _\mu$ and $^*\gamma$ its Hodge dual, and
as usual $\overline{\psi}:=\psi ^\dagger\gamma ^0$ and
\begin{equation}
\overline{D\psi}:=(D\psi )^\dagger\gamma ^0 = d\overline{\psi}
+i\,\overline{\psi}(\Gamma ^{\alpha\beta}\sigma _{\alpha\beta}
+\vartheta ^\mu \pi _\mu )\,.\label{covder2}
\end{equation}

The Dirac matter action (\ref{mattlagrang}) in the presence of
gravity has the peculiarity of including the intrinsic
translational contributions required by the nonlinear gauge
approach to PGT, as seen from the covariant derivatives
(\ref{covder1}) and (\ref{covder2}). However, it is interesting to
notice that these contributions manifest themselves as a mass term
to be added to an explicitly Lorentz-invariant (rather than
Poincar\'e-invariant) Dirac action. Indeed, let us separate the
translational parts of (\ref{covder1}) and (\ref{covder2}) as
\begin{equation}
D\psi =:\widetilde{D}\psi -i\vartheta ^\mu\pi _\mu\psi\,,\qquad
\overline{D\psi}=:\overline{\widetilde{D}\psi}
+i\,\overline{\psi}\,\vartheta ^\mu\pi _\mu\,,\label{covderdec}
\end{equation}
where we denote with tildes the translation-independent pieces
with the standard form of Lorentz covariant derivatives. By
replacing (\ref{covderdec}) in (\ref{mattlagrang}), we realize
that the latter transforms into
\begin{equation}
L_D={i\over2}\,(\overline{\psi}\,\,^*\gamma\wedge\widetilde{D}\psi
+\overline{\widetilde{D}\psi}\wedge\,^*\gamma\psi )
+\,^*m\,\overline{\psi}\psi\,.\label{lagrang2}
\end{equation}
To get (\ref{lagrang2}) we made use of the fact that $\vartheta
^\alpha\wedge\,^*\vartheta _\beta =\delta ^\alpha _\beta\,\eta $,
with $\eta =\,^*1$ as the 4-dimensional volume element, see
Appendix {\bf A}, so that
\begin{equation}
^*\gamma\wedge\vartheta ^\mu \pi _\mu =-\eta\,\gamma ^\mu \pi _\mu
=\,^*m\,(1+\gamma _5 )\,,\label{calcul1}
\end{equation}
and
\begin{equation}
-\vartheta ^\mu \pi _\mu\wedge\,^*\gamma =-\eta\,\pi _\mu \gamma
^\mu =\,^*m\,(1-\gamma _5 )\,.\label{calcul2}
\end{equation}
The particular combination of (\ref{calcul1}) and (\ref{calcul2})
in the matter action cancels out the $\gamma _5$ contribution,
only remaining the background mass term in (\ref{lagrang2}). Thus,
the nonlinear PGT approach to the coupling of translations to
Dirac fields predicts the latter ones to be massive.

\section{\bf Hamiltonian treatment of PGT}

\subsection{\bf Remark on the diversity of equivalent nonlinear approaches}

Previous section was devoted to a nonlinear approach to the gauge
theory of the Poincar\'e group ---namely the one with auxiliary
subgroup $H={\rm Lorentz}$--- useful to support Lagrangian
dynamics of spacetime. However, we recall that given
$G=$Poincar\'e, the choice of $H\subset G$ is not uniquely
predetermined. The outline of Section {\bf II} showed that
nonlinear realizations of a given group $G$ require to fix, in
addition to the total symmetry group $G$ itself, a subgroup
$H\subset G$ enabling the $G$--gauge transformations to act on
representation fields of $H$. No breaking of the original
$G$--symmetry is needed for it to be realized through explicitly
$H$--symmetric quantities. We are free to select any among the
available subgroups $H\subset G$ in order to construct diverse
versions of the gauge theory of one and the same group $G$.
(Notice that the usual gauge theories of internal groups based on
linear realizations rest on the particular NLR corresponding to
the choice $H=G$.) Nonlinear gauge approaches to a group $G$
corresponding to different auxiliary subgroups $H_1$, $H_2$ are
equivalent to each other, being possible to relate them by means
of gauge--like redefinitions of the fields, as we will show in
Section {\bf V}. Thus descriptions of the local group $G$ with
either $H_1$ or $H_2$ constitute different realizations of the
same gauge theory.

In the present section we are going to develop a nonlinear local
realization of the Poincar\'e group having as auxiliary subgroup
$H=SO(3)$ instead of the Lorentz group considered previously. The
resulting $SO(3)$--covariant formalism reveals to be useful for
the Hamiltonian treatment of the PGT approach to gravity. As
before, we characterize dynamical variables by means of
differential forms. An exterior calculus formulation of
Hamiltonian dynamics, fit to gauge theories, is briefly outlined
in the following. It is mainly based on a proposal by Wallner
\cite{Wallnerphd}, suitably adapted to the present nonlinear
approach to PGT with $H=SO(3)$.

\subsection{Poincar\'e invariant foliation of spacetime}

Nonlinear realizations of $G=$Poincar\'e with $H=SO(3)$, as
derived immediately from the general formalism on NLR's of Section
{\bf II}, are displayed in Appendix {\bf C}. The quantities
introduced there will be invoked in what follows in the order
needed for our purposes.

First of all, observe that instead of the four--dimensional
representation (\ref{tetrad}) of the tetrad transforming as a
Lorentz covector as shown in (\ref{varsconns}), in the nonlinear
approach of Appendix {\bf C} we find the coframe $\vartheta
^\alpha$ splitted through definition (\ref{App2.16a}) into an
$SO(3)$ singlet $\hat{\vartheta}^0$ plus an $SO(3)$ covector
$\hat{\vartheta}^a\,$, whose explicit gauge transformations are
given by (\ref{App2.19a}) and (\ref{App2.19b}) respectively. The
invariance (\ref{App2.19a}) of the time component
$\hat{\vartheta}^0$ suggests to perform an invariant foliation of
spacetime into spatial slices as follows.

From the 1--form basis (\ref{App2.16a}), we define the dual vector
basis $\hat{e}_\alpha $ such that
$\hat{e}_\alpha\rfloor\hat{\vartheta} ^\beta =\,\delta _\alpha
^\beta\,$. Starting from the relation $\left[
\hat{e}_\alpha\,,\hat{e}_\beta\,\right] =\,\left(
\hat{e}_\alpha\rfloor \hat{e}_\beta\rfloor d\,\hat{\vartheta}
^\gamma\,\right) \hat{e}_\gamma\,$ holding in the 4--dimensional
space, the necessary and sufficient condition for a foliation into
3--dimensional hypersurfaces normal to $\hat{e}_{_0}$ to exist,
according to the Frobenius' theorem, is that the spatial
restriction of the former formula yields $\left[
\hat{e}_a\,,\hat{e}_b\,\right] =\,\left( \hat{e}_a\rfloor
\hat{e}_b\rfloor d\,\hat{\vartheta} ^c\,\right) \hat{e}_c\,$, not
involving $\hat{e}_{_0}$ in the r.h.s., or equivalently
\begin{equation}
\hat{\vartheta} ^0\wedge d\,\hat{\vartheta} ^0
=\,0\,.\label{Hamilt1}
\end{equation}
Notice that, according to (\ref{App2.19a}), the foliation
condition (\ref{Hamilt1}) is Poincar\'e invariant. The general
solution of (\ref{Hamilt1}) reads
\begin{equation}
\hat{\vartheta} ^0 =\,u^0\,d\,\tau\,.\label{Hamilt2}
\end{equation}
In view of (\ref{Hamilt2}), let us define $u^\alpha :=\,\partial
_\tau\rfloor\hat{\vartheta} ^\alpha$ such that $\partial _\tau
=\,u^\alpha\,\hat{e}_\alpha =\,u^0\,\hat{e}_{_0} +u^a\,\hat{e}_a$,
so that
\begin{equation}
\hat{e}_{_0} =\,{1\over{u^0}}\left(\,\partial _\tau
-u^a\,\hat{e}_a\,\right)\,,\label{Hamilt3}
\end{equation}
satisfying $\hat{e}_{_0}\rfloor \hat{\vartheta} ^0 =1$. In what
follows, we take (\ref{Hamilt3}) as the invariant timelike vector
field defining the foliation direction of spacetime. Accordingly,
it becomes possible to perform the decomposition of any p--form
$\alpha$ \cite{Wallner:1990ng} into a longitudinal and a
transversal part with respect to (\ref{Hamilt3}) as
\begin{equation}
\alpha =\,\hat{\vartheta} ^0\wedge\alpha _{\bot}+\underline{\alpha
}\,,\label{Hamilt4}
\end{equation}
with respective definitions
\begin{equation}
\alpha _{\bot}:=\,\hat{e}_{_0}\rfloor\alpha\,,\qquad
\underline{\alpha }:=\,\hat{e}_{_0}\rfloor\left(\, \hat{\vartheta}
^0\wedge\alpha\,\right)\,.\label{Hamilt5}
\end{equation}
Correspondingly, the foliation of the Hodge dual (\ref{Hamilt4})
reads
\begin{equation}
^*\alpha =\,\left( -1\right) ^p\hat{\vartheta} ^0 \wedge
{}^\#\underline{\alpha }-{}^\#\alpha _{\bot }\,,\label{Hamilt6}
\end{equation}
where the asterisc $*$ stands for the Hodge dual in four
dimensions while $\#$ represents its three--dimensional
restriction, see \cite{Mielke:1990fn} \cite{Mielke:1992te}
\cite{Wallner:1990ng}. On the other hand, the exterior
differential of any p--form decomposes as
\begin{equation}
d\,\alpha =\,\hat{\vartheta}
^0\wedge\left[\,{\it{l}}_{\hat{e}_{_0}}
\underline{\alpha}-{1\over{u^0}}\,\underline{d}\, \left(
\,u^0\,\alpha _{\bot }\right)\,\right] +\underline{d}\,
\underline{\alpha}\,,\label{Hamilt7}
\end{equation}
where we introduced the Lie derivative with respect to
$\hat{e}_{_0}\,$ defined as ${\it{l}}_{\hat{e}_{_0}}\alpha :=
\,d\,\left( \hat{e}_{_0}\rfloor\alpha\,\right) + \left(
\hat{e}_{_0}\rfloor d\,\alpha\,\right)\,$, reducing in particular,
for the transversal part $\underline{\alpha}$ of (\ref{Hamilt4}),
to
\begin{equation}
{\it{l}}_{\hat{e}_{_0}}\underline{\alpha} = \left(
\hat{e}_{_0}\rfloor
d\,\underline{\alpha}\,\right)\,.\label{Hamilt8}
\end{equation}

The present tetrad--adapted spacetime foliation allows a
considerable simplification of the Hamiltonian formalism, mainly
when applied to PGT. The obvious reason is that we do not
distinguish the foliation direction (determined by the time--like
vector upon which the foliation of spacetime is performed) from
the time component (\ref{Hamilt3}) of the vector basis
$\hat{e}_\alpha$ dual to the PGT coframe (\ref{App2.16a}). That
is, the foliation direction is aligned (or even identified) with
the gauge invariant time vector naturally derived from the PGT
approach.

In Appendix {\bf D}, the present foliation procedure is applied to
several quantities defined in Appendix {\bf C} in the context of
the nonlinear realization of the Poincar\'e group with $H=SO(3)$.
In particular, the decomposition into longitudinal and transversal
parts is performed, of the different pieces of torsion and
curvature necessary for the Hamiltonian treatment of PGT to be
developed next.

\subsection{\bf Hamiltonian formalism in terms of differential forms}

Let us outline a Hamiltonian formalism in the language of exterior
calculus \cite{Wallnerphd}, with the foliation procedure exposed
above incorporated to it. We consider a general gauge theory, its
Lagrangian density 4--form
\begin{equation}
L=\,L\left( A\,,d\,A\,\right)\label{Hamilt9}
\end{equation}
depending on the gauge potential 1--form $A$ and on its exterior
derivative $dA\,$. We require the Frobenius foliation condition
(\ref{Hamilt1}) to hold, so that the time component of the tetrad
reduces to (\ref{Hamilt2}). (In practice, this is equivalent to
work in the time gauge, in which only one degree of freedom of
$\hat{\vartheta}^0$ remains different from zero.) Then, in view of
(\ref{Hamilt4}) and (\ref{Hamilt7}), the gauge potential and its
differential decompose into longitudinal and transversal parts as
\begin{equation}
A =\,\hat{\vartheta}^0 A_{\bot} +\underline{A}\,,\qquad d\,A
=\,\hat{\vartheta}^0\wedge\left[\,{\it{l}}_{\hat{e}_{_0}}
\underline{A}-{1\over{u^0}}\,\underline{d}\, \left( u^0\,A_{\bot
}\right)\,\right] +\underline{d}\,
\underline{A}\,,\label{Hamilt10}
\end{equation}
respectively. On the other hand, being the Lagrangian density a
4--form, its transversal part vanishes, decomposing simply as
\begin{equation}
L =\,\hat{\vartheta}^0\wedge L_{\bot}\,.\label{Hamilt11}
\end{equation}
From the Lagrangian normal part $L_{\bot}$ in (\ref{Hamilt11}) we
define the momenta
\begin{equation}
^\#\pi ^{^{A_{\bot}}} :={{\partial L_{\bot}}\over {\partial
({\it{l}}_{\hat{e}_{_0}} A_{\bot})}}\quad\,,\qquad ^\#\pi
^{^{\underline{A}}} :={{\partial L_{\bot}}\over {\partial
({\it{l}}_{\hat{e}_{_0}}\underline{A}\,)}}\,,\label{Hamilt12}
\end{equation}
and with their help we define the Hamiltonian 3--form
\begin{equation}
{\cal{H}}:=\,u^0\,\left[\,\,{\it{l}}_{\hat{e}_{_0}}
A_{\bot}{}^\#\pi ^{^{A_{\bot}}}
+{\it{l}}_{\hat{e}_{_0}}\underline{A} \wedge {}^\#\pi
^{^{\underline{A}}} -L_{\bot}\,\right]\,.\label{Hamilt13}
\end{equation}
In view of (\ref{Hamilt2}), from (\ref{Hamilt13}) we reconstruct
the Lagrangian density (\ref{Hamilt11}) by multiplying by $d\tau$,
getting
\begin{equation}
L =\,d\tau\wedge\left(\,u^0\,{\it{l}}_{\hat{e}_{_0}}
A_{\bot}{}^\#\pi ^{^{A_{\bot}}}
+u^0\,{\it{l}}_{\hat{e}_{_0}}\underline{A}\wedge {}^\#\pi
^{^{\underline{A}}}-{\cal{H}}\,\,\right)\,.\label{Hamilt14}
\end{equation}
Variations of (\ref{Hamilt14}), with ${\cal{H}}$ taken to be a
functional of the gauge potentials and their momenta, yield the
field equations
\begin{equation}
u^0\,{\it{l}}_{\hat{e}_{_0}} A_{\bot}={{\delta
{\cal{H}}}\over{\delta\,{}^\#\pi ^{^{A_{\bot}}}}}\,,\qquad
u^0\,{\it{l}}_{\hat{e}_{_0}} \underline{A} ={{\delta
{\cal{H}}}\over{\delta\,{}^\#\pi ^{^{\underline{A}}}}}\,,\qquad
u^0\,{\it{l}}_{\hat{e}_{_0}}{}^\#\pi ^{^{A_{\bot}}} =-{{\delta
{\cal{H}}}\over{\delta A_{\bot}}}\,,\qquad
u^0\,{\it{l}}_{\hat{e}_{_0}}{}^\#\pi ^{^{\underline{A}}}=-{{\delta
{\cal{H}}}\over{\delta \underline{A}}}\,.\label{Hamilt15}
\end{equation}
(As a technical detail, we follow the convention of putting the
variations of the generalized coordinates to the left and those of
their conjugate momenta to the right.) On the other hand, the Lie
derivative of an arbitrary p--form defined on the 3--space slices
and being a functional of the dynamical variables can be expanded
as
\begin{equation}
{\it{l}}_{\hat{e}_{_0}} \omega =\, {\it{l}}_{\hat{e}_{_0}}
A_{\bot} \,{{\delta \omega }\over{\delta A_{\bot}}}
+{\it{l}}_{\hat{e}_{_0}}\underline{A}\wedge {{\delta \omega }
\over{\delta \underline{A}}} +{{\delta \omega
}\over{\delta\,{}^\#\pi ^{^{A_{\bot}}}}}
\,{\it{l}}_{\hat{e}_{_0}}{}^\#\pi ^{^{A_{\bot}}} +{{\delta \omega
}\over{\delta\,{}^\#\pi ^{^{\underline{A}}}}} \wedge
{\it{l}}_{\hat{e}_{_0}}{}^\#\pi
^{^{\underline{A}}}\,.\label{Hamilt16}
\end{equation}
We define generalized Poisson brackets representing the time
evolution of differential forms as the expressions resulting from
substituting the field equations (\ref{Hamilt15}) into
(\ref{Hamilt16}), that is
\begin{eqnarray}
u^0\,{\it{l}}_{\hat{e}_{_0}} \omega
=\,\left\{\omega\,,{\cal{H}}\,\right\} &:=& {{\delta
{\cal{H}}}\over{\delta\,{}^\#\pi ^{^{A_{\bot}}}}} \,{{\delta
\omega}\over{\delta A_{\bot}}} -{{\delta
\omega}\over{\delta\,{}^\#\pi ^{^{A_{\bot}}}}}\, {{\delta
{\cal{H}}}\over{\delta A_{\bot}}}\nonumber \\
&&+{{\delta {\cal{H}}}\over{\delta\, {}^\#\pi
^{^{\underline{A}}}}}\wedge {{\delta \omega}\over{\delta
\underline{A}}} -{{\delta \omega}\over{\delta\, {}^\#\pi
^{^{\underline{A}}}}} \wedge {{\delta {\cal{H}}}\over{\delta
\underline{A}}}\,.\label{Hamilt17}
\end{eqnarray}
Eq.(\ref{Hamilt17}) is a particular case of the more general
definition
\begin{equation}
\bigl\{\alpha\bigl( x\,\bigr),\, \beta\bigl( y\,\bigr)\bigr\}
:=\int _z \Bigl[{{\partial\,\beta\bigl( y\,\bigr)}
\over{\partial\,{}^\#\Pi _i\bigl( z\,\bigr)}} \wedge
{{\partial\,\alpha\bigl( x\,\bigr)} \over{\partial Q^i\bigl(
z\,\bigr)}} -{{\partial\,\alpha\bigl( x\,\bigr)}
\over{\partial\,{}^\#\Pi _i\bigl( z\,\bigr)}} \wedge
{{\partial\,\beta\bigl( y\,\bigr)} \over{\partial Q^i\bigl(
z\,\bigr)}}\,\Bigr] \wedge\overline{\eta}\bigl( z\,\bigr)
\label{Poisson}
\end{equation}
of Poisson brackets for dynamical variables characterized by
differential forms \cite{Wallnerphd}, where the arbitrary forms
$\alpha $ and $\beta $ are functionals of the canonically
conjugate variables concisely denoted as $Q^i\,$, ${}^\#\Pi _i\,$.
From (\ref{Poisson}) we check that the fundamental Poisson
brackets satisfy
\begin{equation}
\bigl\{ Q^i(x\,)\,,Q^j(y\,)\,\bigr\} =\,0\,,\quad \bigl\{ {}^\#\Pi
_i(x\,)\,,{}^\#\Pi _j(y\,)\,\bigr\} =\,0\,,\label{brackets1}
\end{equation}
\begin{equation}
\bigl\{ Q^i(x\,)\,,{}^\#\Pi _j(y\,)\,\bigr\} =\,\delta
^i_j\,\delta ^3(x-y\,)\,,\label{brackets2}
\end{equation}
as expected. Poisson brackets (\ref{Hamilt17}) provide the formal
instrument needed to calculate the time evolution of dynamical
variables from the Hamiltonian 3--form (\ref{Hamilt13}). Let us
mention a few theorems concerning them, useful for practical
calculations. From definition (\ref{Hamilt17}) follows the
antisymmetry condition
\begin{equation}
\left\{\omega\,,{\cal{H}}\,\right\}
=-\left\{{\cal{H}}\,,\omega\,\right\}\,.\label{Hamilt18}
\end{equation}
In view of the chain rule of the Lie derivative, that is
${\it{l}}_{\hat{e}_{_0}}\left( \sigma\wedge\omega\,\right)
=\,{\it{l}}_{\hat{e}_{_0}}\sigma\wedge\omega +\sigma\wedge
{\it{l}}_{\hat{e}_{_0}}\omega \,$, we deduce the distributive
property
\begin{equation}
\left\{\,\sigma\wedge\omega\,,{\cal{H}}\,\right\}
=\,\left\{\,\sigma\,,{\cal{H}}\,\right\}\wedge\omega
+\sigma\wedge\left\{\,\omega\,,{\cal{H}}\,\right\}\,.\label{Hamilt19}
\end{equation}
From the normal part of the identity $d\wedge d\,\alpha
\equiv\,0$, namely ${\it{l}}_{\hat{e}_{_0}}
\underline{d}\,\underline{\alpha}
-{1\over{u^0}}\,\underline{d}\left( u^0\, {\it{l}}_{\hat{e}_{_0}}
\underline{\alpha }\,\right)\equiv\,0 \,$, it follows
\begin{equation}
\left\{\,\underline{d}\,\underline{\alpha}\,,{\cal{H}}\,\right\}
-\underline{d}\left\{\underline{\alpha}\,,{\cal{H}}\,\right\}
=\,0\,,\label{Hamilt20}
\end{equation}
generalizable to any form defined on the transversal 3--spaces.
With these theorems at hand, we are ready to attack the
Hamiltonian dynamics of a PGT gravitational system.

\subsection{\bf Hamiltonian constraints of PGT}

As in Section {\bf III}, we consider the Einstein--Cartan
Lagrangian 4--form (\ref{EinstCartlagr}), to which we are going to
apply the Hamiltonian formalism outlined previously, completed by
taking into account the fact that PGT--gravity constitutes a
constrained system \cite{Dirac50}, \cite{Dirac64},
\cite{Hanson:1976}, \cite{Blagojevic:1981mm},
\cite{Blagojevic2001}. We make use of the nonlinear version of PGT
of Appendices {\bf C} and {\bf D}. Due to the formal identity of
definitions (\ref{App2.16a}), (\ref{App2.16b}), (\ref{App2.16c})
with gauge transformations, PGT invariants can be alternatively
expressed in terms of quantities corresponding to different NLR's,
so that the PGT invariant action (\ref{EinstCartlagr}) can be
rewritten in terms of the nonlinear variables
$\hat{\Gamma}^{0a}=:\,X^a\,$, $\hat{\Gamma}^{ab}=:\,\epsilon
^{ab}{}_c\,A^c\,$ defined in (\ref{App2.16b}), (\ref{App2.16c}).
The new form we get for (\ref{EinstCartlagr}) reads
\begin{equation}
L=-{1\over{2l^2}}\,\hat{R}_{\alpha\beta}\wedge\hat{\eta}^{\alpha\beta}
+{{\Lambda
}\over{l^2}}\,\hat{\eta}=-{1\over{l^2}}\,\Bigl[\,D\,X^a\wedge\overline{\eta}_a
+\hat{\vartheta}^0\wedge\,\bigl( \hat{\vartheta}_a\wedge
{\cal{R}}^a
-\Lambda\,\overline{\eta}\,\bigr)\,\Bigr]\,.\label{Hamilt22}
\end{equation}
In (\ref{Hamilt22}), the components of the four--dimensional
nonlinear Lorentz curvature relate to the corresponding $SO(3)$
quantities as shown in (\ref{App2.24}), while the $SO(3)$
eta--basis elements are given in (\ref{Hamilt31}). In order to
calculate the momenta as defined in (\ref{Hamilt12}), we have to
find out the normal part of the Lagrangian (\ref{Hamilt22}).
Making use of the decompositions
(\ref{Hamilt26})--(\ref{Hamilt30}) one gets
\begin{equation}
L_{\bot}=-{1\over{l^2}}\,\left\{\,\left[\,{\cal
\L\/}_{e_{_0}}\underline{X}^a -{1\over{u^0}}\,\underline{D}\left(
u^0\,X^a_{\bot}\,\right)\right] \wedge\overline{\eta}_a +\vartheta
^a\wedge\,\underline{\cal{R}}_a
-\Lambda\,\overline{\eta}\,\right\}\,.\label{Hamilt33}
\end{equation}
(In (\ref{Hamilt33}) and from now on we simplify the notation by
suppressing the hat over the $SO(3)$--valued coframes and basis
vectors.) The only nonvanishing momentum obtained from
(\ref{Hamilt33}) is
\begin{equation}
^\#\pi ^{^{\underline{X}}}_a :=\,{{\partial
L_{\bot}}\over{\partial\left(
{\it{l}}_{e_{_0}}\underline{X}^a\,\right)}}
=-{1\over{l^2}}\,\overline{\eta}_a\,,\label{Hamilt34}
\end{equation}
while the remaining ones $^\#\pi ^{u^0}$, $^\#\pi ^{\vartheta}_a
$, $^\#\pi ^{^{A_{\bot}}}_a$, $^\#\pi ^{^{\underline{A}}}_a$ and
$^\#\pi ^{^{X_{\bot}}}_a$ equal zero. All of them together with
(\ref{Hamilt34}) constitute the set of primary constraints.

The total Hamiltonian 3--form of a constrained system is built as
follows. Starting from the canonical Hamiltonian (\ref{Hamilt13})
---adapted in our case to the variables $u^0$, $\vartheta ^a$, $ A_{\bot}^a$,
$\underline{A}^a$, $X_{\bot}^a$, $\underline{X}^a$---, we rewrite
it, whenever possible, in terms of covariant expressions, and then
we replace the factors multiplying the primary constraints by
Lagrange multipliers $\beta ^i\,$. So we get
\begin{eqnarray}
{\cal{H}}&=&\,u^0\,\Bigl\{\,{1\over{l^2}}\left(\,
X^a_{\bot}\,\underline{D}\,\overline{\eta}_a +\vartheta _a
\wedge\underline{\cal{R}}^a -\Lambda\,
\overline{\eta}\,\,\right) \nonumber \\
&&-A_{\bot}^a\left[\,\underline{D}\,{}^\#\pi ^{^{\underline{A}}}_a
+\overline{\eta}_{ab}{}^c\, \left( X_{\bot}^b\,{}^\#\pi
^{^{X_{\bot}}}_c + \underline{X}^b\wedge {}^\#\pi
^{^{\underline{X}}}_c +\vartheta ^b\wedge {}^\#\pi
^{\vartheta}_c\,\right) \,\right]\,\Bigr\}  \nonumber \\
&&+\beta _{u^0}\,^\#\pi ^{u^0} +\beta _{\vartheta}^a\wedge{}^\#\pi
^{\vartheta}_a +\beta _{_{A_{\bot}}}^a\,^\#\pi ^{^{A_{\bot}}}_a
+\beta _{_{\underline{A}}}^a\wedge {}^\#\pi ^{^{\underline{A}}}_a
+\beta _{_{X_{\bot}}}^a\,^\#\pi ^{^{X_{\bot}}}_a  \nonumber \\
&&+\beta _{_{\underline{X}}}^a\wedge\Bigl(\,^\#\pi
^{^{\underline{X}}}_a
+{1\over{l^2}}\,\overline{\eta}_a\,\Bigr)\,.\label{Hamilt35}
\end{eqnarray}
From (\ref{Hamilt35}), the time evolution of any dynamical
variable is calculable in principle with the help of Poisson
brackets of the form (\ref{Hamilt17}), suitably generalized to the
whole set of conjugate variables $u^0\,,{}^\#\pi ^{u^0}$;
$\vartheta ^a\,,{}^\#\pi ^{\vartheta}_a$; $A_{\bot}^a\,,{}^\#\pi
^{^{A_{\bot}}}_a$; $\underline{A}^a\,,{}^\#\pi
^{^{\underline{A}}}_a$; $X_{\bot}^a\,,{}^\#\pi ^{^{X_{\bot}}}_a$;
$\underline{X}^a\,,{}^\#\pi ^{^{\underline{X}}}_a\,$.

Primary constraints are required to be stable. That is, their
respective evolutions in time are enforced to vanish, giving rise
to four secondary constraints plus two conditions on the Lagrange
multipliers as follows. On the one hand, the evolution equations
\begin{eqnarray}
u^0\,{\it{l}}_{e_{_0}}\,{}^\#\pi ^{u^0}
&=&-{1\over{l^2}}\,\left(\,\varphi ^{^{(0)}} + X^a_{\bot}\,\varphi
^{^{(3)}}_a\,\right) +A_{\bot}^a\,\varphi ^{^{(1)}}_a\,,\label{Hamilt36a} \\
u^0\,{\it{l}}_{e_{_0}}\,{}^\#\pi ^{^{A_{\bot}}}_a &=&\hskip0.4cm
u^0\,\varphi ^{^{(1)}}_a\,,\label{Hamilt36b} \\
u^0\,{\cal \L\/}_{e_{_0}}\,{}^\#\pi ^{^{\underline{A}}}_a
&=&-{1\over{l^2}}\,\varphi ^{^{(2)}}_a\,,\label{Hamilt36c} \\
u^0\,{\cal \L\/}_{e_{_0}}\,{}^\#\pi ^{^{X_{\bot}}}_a
&=&-{{u^0}\over{l^2}}\,\varphi ^{^{(3)}}_a\,,\label{Hamilt36d}
\end{eqnarray}
when put equal to zero, yield the secondary constraints
\begin{eqnarray}
\varphi ^{^{(0)}}&:=&\,\vartheta _a \wedge\underline{\cal{R}}^a
-\Lambda\,\overline{\eta}\,,\label{Hamilt37a} \\
\varphi ^{^{(1)}}_a&:=&\,\underline{D}\,{}^\#\pi
^{^{\underline{A}}}_a +\epsilon _{ab}{}^c\left(
X_{\bot}^b\,{}^\#\pi ^{^{X_{\bot}}}_c + \underline{X}^b\wedge
{}^\#\pi ^{^{\underline{X}}}_c +\vartheta ^b\wedge {}^\#\pi
^{\vartheta}_c\,\right)\,,\label{Hamilt37b} \\
\varphi ^{^{(2)}}_a &:=&\,\underline{D} (u^0\, \vartheta _a)
+u^0\,X_{\bot}^b\vartheta _b\wedge \vartheta _a\,,\label{Hamilt37c} \\
\varphi ^{^{(3)}}_a&:=&\,\underline{D}
\,\overline{\eta}_a\,,\label{Hamilt37d}
\end{eqnarray}
while on the other hand the vanishing of the time evolution of the
remaining primary constraints, namely
\begin{eqnarray}
&&u^0\,{\cal \L\/}_{e_{_0}}\,\Bigl(\,^\#\pi ^{^{\underline{X}}}_a
+{1\over{l^2}}\,\overline{\eta}_a\,\Bigr)
=-{1\over{l^2}}\,\overline{\eta}_{ab}\wedge\left(\,\beta
_{\vartheta}^b +u^0\,\underline{X}^b\,\right)\,, \label{Hamilt38a}\\
&&u^0\,{\cal \L\/}_{e_{_0}}\,{}^\#\pi ^{\vartheta}_a
=-{1\over{l^2}}\,\Bigl\{\,\Bigl[\,\beta^b_{_{\underline{X}}}
-\underline{D}\left( u^0\,X^b_{\bot}\,\right)\,\Bigr]
\wedge\overline{\eta}_{ab} +u^0\,\Bigl(\, \underline{\cal{R}}_a
-\Lambda\,\overline{\eta}_a \,\Bigr)\,\Bigr\}\,,\label{Hamilt38b}
\end{eqnarray}
fixes conditions on the Lagrange multipliers $\beta_{\vartheta}^a$
and $\beta _{_{\underline{X}}}^a$ respectively. By solving
(\ref{Hamilt38a}) and (\ref{Hamilt38b}) equaled to zero we get
\begin{equation}
\beta_{\vartheta}^a =-u^0\,\underline{X}^a\,,\label{Hamilt39}
\end{equation}
as much as
\begin{eqnarray}
\beta_{_{\underline{X}}}^a&=&\,\underline{D}\left(
u^0\,X^a_{\bot}\,\right) +u^0\,\left[\,{}^\#\underline{\cal{R}}^a
-e^a\rfloor\left(\,\vartheta
_b\wedge{}^\#\underline{\cal{R}}^b\,\right) -{1\over2}\,\vartheta
^a\,{}^\#\left(\vartheta _b\wedge\underline{\cal{R}}^b\,\right)
+{{\Lambda}\over2}\,\vartheta ^a\,\right]\nonumber \\
&\approx &\,\underline{D}\left( u^0\,X^a_{\bot}\,\right)
+u^0\,\left[\,{}^\#\underline{\cal{R}}^a
-e^a\rfloor\left(\,\vartheta
_b\wedge{}^\#\underline{\cal{R}}^b\,\right)\,\right]\,.\label{Hamilt40}
\end{eqnarray}
The simplification in (\ref{Hamilt40}) follows from taking
$\varphi ^{^{(0)}}$ in (\ref{Hamilt37a}) into account. (The symbol
$\approx $ indicates that the equation holds weakly, that is in
the subspace of the phase space where all constraints hold.)

Let us deduce several consequences of the secondary constraints
(\ref{Hamilt37b})--(\ref{Hamilt37d}). Having in mind also the
primary ones, the constraint (\ref{Hamilt37b}) reduces weakly to
$\varphi ^{^{(1)}}_a\approx -{1\over{l^2}}\, \vartheta
_a\wedge\vartheta _b\wedge\underline{X}^b\approx 0\,$ implying
$\vartheta _a\wedge\underline{X}^a\approx 0\,$. On the other hand,
the vanishing of (\ref{Hamilt37d}) implies $\varphi ^{^{(3)}}_a
:=\underline{D}\,\overline{\eta}_a =\,
\overline{\eta}_{ab}\wedge\underline{D}\,\vartheta ^b
\approx\,0\,$. Replacing here $\underline{D}\,\vartheta ^b\approx
-\left(\,\underline{d}\log u^0 +X^a_{\bot}\,\vartheta
_a\,\right)\wedge\vartheta ^b$ as deduced from the constraint
(\ref{Hamilt37c}), one gets $\left(\,\underline{d}\log u^0
+X^b_{\bot}\,\vartheta
_b\,\right)\wedge\overline{\eta}_a\approx\,0\,$, proving that
$\underline{d}\, \log u^0 + X^a_{\bot}\,\vartheta _a \approx\,0$
is a constraint by itself, so that $\underline{D}\,\vartheta ^a$,
being proportional to it, also vanishes weakly. In summary,
(\ref{Hamilt37b})--(\ref{Hamilt37d}) with the help of the primary
constraints give rise to
\begin{equation}
\underline{d}\,\log u^0 +X_{\bot}^a\,\vartheta _a\approx
0\,,\qquad \vartheta _a\wedge\underline{X}^a\approx 0\,,\qquad
\underline{D}\,\vartheta ^a\approx 0\,,\label{Hamilt41}
\end{equation}
whose geometrical meaning as the vanishing of several torsion
pieces becomes clear by comparison with (\ref{torsfoliat1}),
(\ref{torsfoliat2}).

On the other hand, (\ref{Hamilt40}) can be further simplified. In
view of the definition of $\underline{\cal{R}}^a$ in
(\ref{Hamilt30}), we put $\vartheta
_a\wedge{}^\#\underline{\cal{R}}^a =\vartheta
_a\wedge{}^\#\underline{F}^a  +{1\over 2}\,\left[ e_a\rfloor
e_b\rfloor\left(\vartheta _d\wedge
\underline{X}^d\right)\,\right]\epsilon
^{ab}{}_c\,{}^\#\underline{X}^c\,$, where the last term vanishes
according to the second equation in (\ref{Hamilt41}), and so does
the first term of the r.h.s. since the covariant differential of
the third equation in (\ref{Hamilt41}) yields
$\underline{D}\,\underline{D}\,\vartheta ^a
=\overline{\eta}^a{}_b\wedge\underline{F}^b =\vartheta
^a\wedge\vartheta _b\wedge\,{}^\#\underline{F}^b\approx 0\,$, so
that $\vartheta _a\wedge\,{}^\#\underline{F}^a\approx 0\,$. Thus
we conclude
\begin{equation}
\vartheta _a\wedge{}^\#\underline{\cal{R}}^a =0\,.\label{Hamilt42}
\end{equation}
Replacing (\ref{Hamilt42}) in (\ref{Hamilt40}), the latter reduces
to its ultimate form
\begin{equation}
\beta_{_{\underline{X}}}^a\approx \,\underline{D}\left(
u^0\,X^a_{\bot}\,\right)
+u^0\,{}^\#\underline{\cal{R}}^a\,.\label{Hamilt43}
\end{equation}

According to the general treatment of constrained systems
\cite{Dirac50} \cite{Dirac64} \cite{Hanson:1976}, we furthermore
have to require the secondary constraints
(\ref{Hamilt37a})--(\ref{Hamilt37d}) to be stable in time. The
time evolution of (\ref{Hamilt37b}) is found to automatically
satisfy
\begin{equation}
u^0\,{\cal \L\/}_{e_{_0}}\,\varphi ^{(1)}_a\approx
\,0\,,\label{Hamilt44}
\end{equation}
by simply taking into account the already known constraints.
However, new conditions on the Lagrange multipliers are necessary
to guarantee the stability of (\ref{Hamilt37c}) and
(\ref{Hamilt37d}). For the latter we find
\begin{equation}
u^0\,{\cal \L\/}_{e_{_0}}\,\varphi ^{(3)}_a
\approx\overline{\eta}_{ab}\wedge\left(\,\underline{D}\,\beta
_{\vartheta}^b -\overline{\eta}^b{}_c\wedge\beta
_{_{\underline{A}}}^c\,\right)\,,\label{Hamilt45}
\end{equation}
implying, when enforced to vanish,
\begin{equation}
\beta _{_{\underline{A}}}^a ={}^\#\underline{D}\,\beta
_{\vartheta}^a -e^a\rfloor\left(\,\vartheta
_b\wedge{}^\#\underline{D}\,\beta _{\vartheta}^b\,\right)
-{1\over2}\,\vartheta ^a\,{}^\#\left(\vartheta _b\wedge
\underline{D}\,\beta _{\vartheta}^b\,\right)\,,\label{Hamilt46}
\end{equation}
which in view of (\ref{Hamilt39}) with (\ref{Hamilt41}) reduces to
\begin{equation}
\beta _{_{\underline{A}}}^a \approx u^0\,\left[\,\epsilon
^a{}_{bc}\,X_{\bot}^b\underline{X}^c
-\,{}^\#\underline{D}\,{\underline{X}}^a
+e^a\rfloor\left(\,\vartheta
_b\wedge{}^\#\underline{D}\,{\underline{X}}^b\,\right)\,\right]\,.\label{Hamilt47}
\end{equation}
On the other hand, time evolution of (\ref{Hamilt37c}) is
calculated to be
\begin{equation}
u^0\,{\cal \L\/}_{e_{_0}}\,\varphi ^{(2)}_a\approx -u^0\,\vartheta
_a\wedge\left[\,\underline{d}\, \left({{\beta
_{u^0}}\over{u^0}}\right) +\beta _{_{X_\bot}}^b\,\vartheta _b
+X_{\bot}^b\,\beta ^{\vartheta}_b\,\,\right]\,,\label{Hamilt48}
\end{equation}
whose vanishing ensures the stability ${\cal
\L\/}_{e_{_0}}\left(\,\underline{d} \log u^0
+X_{\bot}^a\,\vartheta _a\,\right)\approx\,0\,$ of the first
equation in (\ref{Hamilt41}). Finally we require the stability of
the constraint (\ref{Hamilt37a}). Making use of (\ref{Hamilt39})
and (\ref{Hamilt43}) we find
\begin{equation}
u^0\,{\it{l}}_{e_{_0}}\,\varphi ^{(0)}\approx
-{1\over{u^0}}\,\underline{d}\left[\,u^0\left(\,\vartheta
_a\wedge\beta _{_{\underline{A}}}^a -u^0
X_{\bot}^a\,\overline{\eta}_{ab}\wedge\underline{X}^b\,\right)\,\right]\,.\label{Hamilt49}
\end{equation}
The differentiated quantity in (\ref{Hamilt49}) can be calculated
in view of (\ref{Hamilt47}) with (\ref{Hamilt41}), yielding
\begin{equation}
\vartheta _a\wedge\beta _{_{\underline{A}}}^a -u^0
X_{\bot}^a\,\overline{\eta}_{ab}\wedge\underline{X}^b \approx
u^0\,{}^\#\underline{D}\,\underline{X}^a + u^0
X_{\bot}^a\,{}^\#\left[ e_a\rfloor\left(\vartheta
_b\wedge\underline{X}^b\,\right)\,\right]\approx
u^0\,{}^\#\underline{D}\,\underline{X}^a\,,\label{Hamilt50}
\end{equation}
so that (\ref{Hamilt49}) transforms into
\begin{equation}
u^0\,{\it{l}}_{e_{_0}}\,\varphi ^{(0)}\approx
-{1\over{u^0}}\,\underline{d}\left[\,(u^0)^2\,\varphi
^{(4)}\,\right]\,,\label{Hamilt51}
\end{equation}
where we defined
\begin{equation}
\varphi ^{(4)} :=\,\vartheta _a\wedge
{}^\#\underline{D}\,\underline{X}^a \,.\label{Hamilt52}
\end{equation}
The vanishing of (\ref{Hamilt51}) is the stability condition of
(\ref{Hamilt37a}), so that in principle (\ref{Hamilt51}) should be
taken as a new constraint. However, one can check that for
(\ref{Hamilt51}) to be stable, $\varphi ^{(4)}$ as defined in
(\ref{Hamilt52}) must vanish, so that (\ref{Hamilt52}) itself,
rather than the less restrictive condition (\ref{Hamilt51}), is to
be considered as the new constraint.

In view of the vanishing of (\ref{Hamilt52}), the Lagrange
multiplier (\ref{Hamilt47}) reduces to
\begin{equation}
\beta _{_{\underline{A}}}^a \approx u^0\,\left(\,\epsilon
^a{}_{bc}\,X_{\bot}^b\underline{X}^c
-\,{}^\#\underline{D}\,{\underline{X}}^a\,\right)\,,\label{Hamilt53}
\end{equation}
and finally one can prove that the constraint (\ref{Hamilt52}) is
stable, thus completing our search for the constraints (and for
the solved Lagrange multipliers) of the theory.

\subsection{\bf Comparison to the Lagrangian approach to PGT}

The meaning of the gravitational equations obtained in the context
of the Hamiltonian approach to PGT becomes clarified by comparing
them with the ordinary PGT Lagrangian equations (\ref{Einsteqn}),
(\ref{zerotorsionbis}) derived from the same action
(\ref{EinstCartlagr}). Besides the constraints of previous
subsection, we have to consider the evolution in time of the
spatial triad $\vartheta ^a$, of the $SO(3)$ connection
$\underline{A}^a$ and of the nonlinear boost connection
$\underline{X}^a$. As we know, their Lie derivatives along the
time direction $e_{_0}$ are found from (\ref{Hamilt35}) with the
help of the Poisson brackets (\ref{Hamilt17}). Taking into account
the previously calculated values of the Lagrange multipliers
(\ref{Hamilt39}), (\ref{Hamilt43}) and (\ref{Hamilt53}), and the
definitions of covariantized Lie derivatives in
(\ref{torsfolaux}), (\ref{Hamilt29}) and (\ref{Hamilt27a}), we get
\begin{eqnarray}
u^0\,{\cal \L\/}_{e_{_0}}\vartheta ^a =\,\beta
_{\vartheta}^a\approx -u^0\,\underline{X}^a\Rightarrow &&{\cal
\L\/}_{e_{_0}}\vartheta ^a +\underline{X}^a\approx 0\,,\label{Hamilt55a} \\
u^0\,F_{\bot}^a =\,\beta _{_{\underline{A}}}^a \approx
u^0\,\left(\,\epsilon ^a{}_{bc}\,X_{\bot}^b\underline{X}^c
-\,{}^\#\underline{D}\,{\underline{X}}^a\,\right)\Rightarrow
&&{\cal{R}}_{\bot}^a \approx
-\,^\#\left(\underline{D}\,\underline{X}^a\right)\,,\label{Hamilt55b} \\
u^0\,{\cal \L\/}_{e_{_0}} \underline{X}^a
=\,\beta_{_{\underline{X}}}^a\approx \,\underline{D}\left(
u^0\,X^a_{\bot}\,\right)
+u^0\,{}^\#\underline{\cal{R}}^a\Rightarrow &&{\cal \L\/}_{e_{_0}}
\underline{X}^a -{1\over{u^0}}\,\underline{D}\left(
u^0\,X^a_{\bot}\,\right)\approx
\,{}^\#\underline{\cal{R}}^a\,.\hskip1.0cm\label{Hamilt55c}
\end{eqnarray}
The evolution equations (\ref{Hamilt55a})--(\ref{Hamilt55c}) plus
the set of constraints found above summarize the PGT Hamiltonian
dynamics we want to compare with the Lagrangian field equations
(\ref{Einsteqn}), (\ref{zerotorsionbis}). In order to do so, we
decompose the latter ones into their longitudinal and transversal
parts making use of the results of Appendix {\bf D}. Let us begin
with the Lagrangian result (\ref{zerotorsionbis}) of vanishing
torsion, which in view of (\ref{torsfoliat1}),
(\ref{torsfoliat2}), takes the form
\begin{eqnarray}
0&=&\,\hat{T}^0
=-\hat{\vartheta}^0\wedge\left(\,\underline{d}\,\log u^0
+X^a_{\bot}\hat{\vartheta}_a\,\right)+\hat{\vartheta}_a\wedge
\underline{X}^a\,,\label{Hamilt57a} \\
0&=&\,\hat{T}^a =\,\hat{\vartheta}^0\wedge\left( {\cal
\L\/}_{\hat{e}_{_0}}\hat{\vartheta}^a +\underline{X}^a\,\right)
+\underline{D}\,\hat{\vartheta}^a\,,\label{Hamilt57b}
\end{eqnarray}
where we reintroduced the hat notation of Appendices {\bf C} and
{\bf D} in order to avoid confusions. It is easy to check that the
four equations contained in (\ref{Hamilt57a}), (\ref{Hamilt57b})
coincide with the Hamiltonian constraints (\ref{Hamilt41})
together with the evolution equation (\ref{Hamilt55a}). The
Hamiltonian equations involved acquire in this way an explicit
geometrical meaning. On the other hand, the Einstein equations
(\ref{Einsteqn}) decompose as the time component
\begin{equation}
0=\,{1\over2}\,\hat{\eta}_{0\beta\gamma}\wedge\hat{R}^{\beta\gamma}
-\Lambda\,\hat{\eta}_0
=-\hat{\vartheta}^0\wedge\left(\hat{\vartheta}_a\wedge
{\cal{R}}_{\bot}^a\,\right) +\left(\,\hat{\vartheta}_a\wedge
\underline{\cal{R}}^a
-\Lambda\,\overline{\eta}\,\,\right)\,,\label{Hamilt58}
\end{equation}
and the spatial components
\begin{equation}
0=\,{1\over2}\,\hat{\eta}_{a\beta\gamma}\wedge\hat{R}^{\beta\gamma}
-\Lambda\,\hat{\eta}_a =-\hat{\vartheta}^0\wedge \Biggl\{\Bigl[\,
{\cal \L\/}_{e_{_0}}\,\underline{X}^b
-{1\over{u^0}}\,\underline{D}\left( u^0\,X^b_{\bot}\,\right)
\,\Bigr]\wedge\overline{\eta}_{ab} +\underline{\cal{R}}_a
-\Lambda\,\overline{\eta}_a\,\Biggr\} -\overline{\eta}_{ab}
\wedge\underline{D}\,\underline{X}^b\,,\label{Hamilt59}
\end{equation}
respectively. Regarding (\ref{Hamilt58}), notice that the
transversal part is the constraint (\ref{Hamilt37a}), while the
vanishing of the longitudinal part follows from (\ref{Hamilt55b})
with the constraint (\ref{Hamilt52}). Furthermore,
(\ref{Hamilt52}) also yields the vanishing of the transversal part
of (\ref{Hamilt59}) since trivially $0\approx
\hat{\vartheta}_a\wedge\varphi
^{^{(4)}}=\hat{\vartheta}_a\wedge\hat{\vartheta}_b\wedge
{}^\#\underline{D}\,\underline{X}^b
=\overline{\eta}_{ab}\wedge\underline{D}\,\underline{X}^b\,$,
where we made use of the Hodge dual relations of Appendix {\bf A},
suitably adapted to three--dimensional space. The vanishing of the
longitudinal part of (\ref{Hamilt59}) follows from performing the
exterior product of (\ref{Hamilt55c}) by $\overline{\eta}_{ab}$.
The coincidence of the resulting expression with the one in
(\ref{Hamilt59}) can be easily proved as follows. We start with
the three--dimensional-adapted identity ${}^\#\underline{\cal
R}^b\wedge\overline{\eta}_a\equiv\hat{\vartheta}_a\wedge
\underline{\cal R}^b$, see Appendix {\bf A}, and contract it with
$\hat{e}_b$. Then, by invoking the constraints (\ref{Hamilt37a})
and (\ref{Hamilt42}) together with the identity
$^\#(\underline{\alpha}\wedge\hat{\vartheta}_b)\equiv
\,(\hat{e}_b\rfloor\,^\#\underline{\alpha} )$, see Appendix {\bf
A}, we get ${}^\#\underline{{\cal
R}}^b\wedge\overline{\eta}_{ab}\approx -(\underline{\cal R}_a
-\Lambda\overline{\eta}_a\,)$. Thus we were able to deduce the
complete set of Lagrangian equations
(\ref{Hamilt57a})--(\ref{Hamilt59}) from the Hamiltonian approach
to PGT.

Observe that the reciprocal derivation is not possible. Indeed,
the longitudinal part of (\ref{Hamilt58}), that is
$\hat{\vartheta}_a\wedge {\cal{R}}_{\bot}^a\approx 0$, is obtained
as the trace of (\ref{Hamilt55b}) provided (\ref{Hamilt52})
vanishes. However, equation (\ref{Hamilt55b}) itself has not a
Lagrangian equivalent. Something similar can be said about
(\ref{Hamilt55c}) and the longitudinal part of (\ref{Hamilt59}).
It is precisely the presence of (\ref{Hamilt55b}) and
(\ref{Hamilt55c}) that makes it possible to put the Hamiltonian
Einstein equations together into a very simple $SO(3)$--covariant
formulation on four--dimensional spacetime. Taking into account
(\ref{Hamilt26}) and (\ref{Hamilt28}), both (\ref{Hamilt55c}) and
the Hodge dual of (\ref{Hamilt55b}) rearrange into
\begin{equation}
D\,X^a -{}^*{\cal{R}}^a\approx\,0\,,\label{Hamilt60}
\end{equation}
while the trace $\hat{\vartheta}_a\wedge {\cal{R}}_{\bot}^a\approx
0$ of (\ref{Hamilt55b}) besides the constraint (\ref{Hamilt37a})
are summarized by the four--dimensional formula
\begin{equation}
\hat{\vartheta}_a\wedge {\cal{R}}^a -\Lambda\,\hat{\eta}_{_0}
\approx\,0\,,\label{Hamilt61}
\end{equation}
being $\hat{\eta}_{_0}:=\hat{e}_{_0}\rfloor\eta
=:\overline{\eta}\,$, see (\ref{Hamilt31}). Equations
(\ref{Hamilt60}), (\ref{Hamilt61}) with the additional conditions
(\ref{Hamilt57a}), (\ref{Hamilt57b}) of vanishing torsion
constitute the condensed form of the Hamiltonian PGT equations
derived from the ordinary Einstein--Cartan action
(\ref{EinstCartlagr}) in vacuum.

\subsection{\bf Relation to Ashtekar variables}

We are interested in calling attention on the close relationship
in which the variables (\ref{App2.16a})--(\ref{App2.16c})
introduced by us in the context of Hamiltonian PGT stand to the
Ashtekar variables, see \cite{Ashtekar:1986yd}
\cite{Ashtekar:1987gu} \cite{Jacobson:1988yy}
\cite{Wallner:1990ng} \cite{Wallner:1992pj}. To make the link
apparent, notice that the Lagrangian dynamics of the
Einstein--Cartan action (\ref{EinstCartlagr}) is not modified by
adding to it a term as
\begin{equation}
L=L_{EC}-\beta\,{1\over{2l^2}}\,\hat{R}^*_{\alpha\beta}\wedge\hat{\eta}^{\alpha\beta}
=-{1\over{2l^2}}\,\left(\hat{R}_{\alpha\beta}
+\beta\,\hat{R}^*_{\alpha\beta}\right)\wedge\hat{\eta}^{\alpha\beta}
+{{\Lambda }\over{l^2}}\,\hat{\eta}\,,\label{Ashtek4a}
\end{equation}
the latter being proportional (with arbitrary constant coefficient
$\beta\,$) to the Lie dual of the curvature, that is to
\begin{equation}
\hat{R}^*_{\alpha\beta}:={1\over
2}\,\hat{\eta}_{\alpha\beta}{}^{\mu\nu}\,\hat{R}_{\mu\nu}\,,\label{Ashtek3}
\end{equation}
(not to be confused with the Hodge dual considered in Appendix
{\bf A}). Indeed, as compared with the Lagrangian approach of
Subsection {\bf III. B}, the contributions of the additional term
in (\ref{Ashtek4a}) to the field equation analogous to
(\ref{zerotorsion}) still imply zero torsion
(\ref{zerotorsionbis}), while the modification of the Einstein
equation (\ref{Einsteqn}) as deduced from (\ref{Ashtek4a}) is
enlarged by a term proportional to
$\hat{R}_\mu{}^\nu\wedge\hat{\vartheta}_\nu\equiv
-\hat{D}\hat{T}_\mu\,$, which also vanishes for vanishing torsion.
So the Lagrangian dynamics derived from (\ref{Ashtek4a}) is
indistinguishable from the one obtained from
(\ref{EinstCartlagr}). We will show that also the Hamiltonian
equations coincide with those of the standard Einstein--Cartan
case.

Let us start by reexpressing (\ref{Ashtek4a}) in terms of the PGT
variables (\ref{App2.16a})--(\ref{App2.16c}) making use of
(\ref{App2.24}), so that the main term in (\ref{Ashtek4a}) becomes
\begin{equation}
{1\over{2}}\,\left(\hat{R}_{\alpha\beta}
+\beta\,\hat{R}^*_{\alpha\beta}\right)\wedge\hat{\eta}^{\alpha\beta}=
\left(\,\vartheta ^0\wedge\vartheta ^a +\beta\,\eta
^{0a}\,\right)\wedge {\cal R}_a + \left(\,\beta\,\vartheta
^0\wedge\vartheta ^a -\eta ^{0a}\,\right)\wedge D\,X_a
\,.\label{betaconst1}
\end{equation}
Then we combine the PGT connection fields (\ref{App2.16b}),
(\ref{App2.16c}), namely the $SO(3)$ connections $A^a$ and the
nonlinear boost connections $X^a$, into a modified $SO(3)$
connection
\begin{equation}
\tilde{A}^a :=A^a +\beta\,X^a\,,\label{Ashtekbeta}
\end{equation}
which we claim to be a variable of the Ashtekar type, as will be
justified by the following development. For later convenience, in
(\ref{Ashtekbeta}) the constant $\beta$ is chosen to be the same
as in (\ref{Ashtek4a}), without further determining its value
\cite{Barbero:1995ap}, non even prejudging for the moment about
its real or complex character. The $SO(3)$ field strength built
from (\ref{Ashtekbeta}) reads
\begin{equation}
\tilde{F}^a :=\,d\,\tilde{A}^a +{1\over2}\,\epsilon
^a{}_{bc}\,\tilde{A}^b\wedge\tilde{A}^c ={\cal R}^a +\beta\,D\,X^a
+(\,\beta ^2 +1\,)\,{1\over2}\,\epsilon ^a{}_{bc}\,X^b\wedge
X^c\,,\label{betafieldstr}
\end{equation}
compare with (\ref{App2.22}), (\ref{App2.23}). Replacing
(\ref{betafieldstr}) in (\ref{betaconst1}) (as much as the
covariant derivative $D X^a$ in (\ref{betafieldstr}), defined as
(\ref{App2.22}), by $\tilde{D} X^a$ in terms of
(\ref{Ashtekbeta})), we get
\begin{equation}
{1\over{2}}\,\left(\hat{R}_{\alpha\beta}
+\beta\,\hat{R}^*_{\alpha\beta}\right)\wedge\hat{\eta}^{\alpha\beta}=
\left(\,\vartheta ^0\wedge\vartheta ^a +\beta\,\eta
^{0a}\,\right)\wedge\tilde{F}_a -(\,\beta ^2 +1\,)\,\bigl[\,\eta
^{0a}\,\wedge\tilde{D}\,X_a +\left(\,\vartheta ^0\wedge\vartheta
^a -\beta\,\eta ^{0a}\,\right)\wedge\,{1\over2}\,\epsilon
_{abc}\,X^b\wedge X^c\bigr]\,.\label{betaconst2}
\end{equation}
We are free to maintain the value of $\beta$ arbitrary or even to
choose it to be real despite the Lorentzian signature we are
dieling with \cite{Barbero:1995ap}, but it is obvious that a major
simplification of (\ref{betaconst2}) follows from taking $\beta ^2
=-1\,$. In particular we fix $\beta =i\,$, so that
(\ref{Ashtek4a}) becomes an action of the Jacobson--Smolin type
\cite{Jacobson:1988yy}, namely
\begin{equation}
L=-{1\over{l^2}}\,^{(-\,)}\hat{R}_{\alpha\beta}\wedge\hat{\eta}^{\alpha\beta}
+{{\Lambda }\over{l^2}}\,\hat{\eta}
=-{1\over{l^2}}\,\left(\,\vartheta ^0\wedge\vartheta ^a +i\,\eta
^{0a}\,\right)\wedge\tilde{F}_a +{{\Lambda
}\over{l^2}}\,\hat{\eta}\,,\label{JacobSmol}
\end{equation}
with the anti--self--dual curvature defined from the curvature and
its Lie dual (\ref{Ashtek3}) as
\begin{equation}
^{(-\,)}\hat{R}_{\alpha\beta}:= {1\over
2}\,\left(\hat{R}_{\alpha\beta}
+i\,\hat{R}^*_{\alpha\beta}\right)\,,\label{antiselfdualcurv}
\end{equation}
thus satisfying ${}^{(-\,)}\hat{R}^*_{\alpha\beta}
=-i\,{}^{(-\,)}\hat{R}_{\alpha\beta}\,$. The components of the
corresponding anti--self--dual connection
\begin{equation}
^{(-\,)}\hat{\Gamma}_{\alpha\beta}:= {1\over
2}\,\left(\hat{\Gamma}_{\alpha\beta}
+i\,\hat{\Gamma}^*_{\alpha\beta}\right)\,,\quad {\rm with}\quad
\hat{\Gamma}^*_{\alpha\beta}:={1\over
2}\,\hat{\eta}_{\alpha\beta}{}^{\mu\nu}\,\hat{\Gamma}_{\mu\nu}
\,,\label{antiselfdualconn}
\end{equation}
relate to the complex Ashtekar connection (\ref{Ashtekbeta}) with
$\beta =i$ as
\begin{equation}
\tilde{A}^a :=A^a +i\,X^a = \epsilon ^a{}
_{bc}\,{}^{(-\,)}\hat{\Gamma}^{bc} =
2i\,{}^{(-\,)}\hat{\Gamma}^{0a}\,,\label{Ashtek1}
\end{equation}
while the field strength (\ref{betafieldstr}) reduces to
\begin{equation}
\tilde{F}^a ={\cal R}^a +i\,D\,X^a\,.\label{Ashtek2b}
\end{equation}

For what follows, it is convenient to rewrite (\ref{JacobSmol})
making use of the identity $\vartheta ^0\wedge\vartheta
^a\wedge\tilde{F}_a \equiv {}^*\tilde{F}_a\wedge\,\eta ^{0a}\,$,
see Appendix {\bf A}, together with the notation of
(\ref{Hamilt31}), as the complex Lagrangian
\begin{equation}
L={1\over{l^2}}\,\Bigl[\,\left(
i\tilde{F}^a-{}^*\tilde{F}^a\,\right)\wedge\overline{\eta}_a
+\Lambda\,\hat{\vartheta}^0\wedge\overline{\eta}\,\,\Bigr]\,,\label{Ashtek4b}
\end{equation}
to which we proceed to apply the Hamiltonian treatment developed
previously. The normal part of (\ref{Ashtek4b}) reads
\begin{equation}
L_{\bot}={1\over{l^2}}\,\left[\,\left(\,i\,\tilde{F}_{\bot}^a
-{}^\#\tilde{\underline{F}}^a\,\right)\wedge\overline{\eta}_a
+\Lambda\,\overline{\eta}\,\,\right]\,,\label{Ashtek5}
\end{equation}
where we used definitions
\begin{equation}
\tilde{F}_{\bot}^a :=\,{\it{l}}_{e_{_0}}\,\underline{\tilde{A}}^a
-{1\over{u^0}}\,\underline{\tilde{D}}\left(\,u^0\,
\tilde{A}_{\bot}^a\,\right)\,,\qquad \underline{\tilde{F}}^a
:=\,\underline{d}\,\underline{\tilde{A}}^a +{1\over2}\,\epsilon
^a{}_{bc}\,\underline{\tilde{A}}^b\wedge \underline{\tilde{A}}^c
\,,\label{Ashtek6}
\end{equation}
analogous to those in (\ref{Hamilt29}). All momenta calculated
from (\ref{Ashtek5}) result to be primary constraints, being the
only nonvanishing one
\begin{equation}
^\#\pi ^{^{\underline{\tilde{A}}}}_a :=\,{{\partial
L_{\bot}}\over{\partial\left(
{\it{l}}_{e_{_0}}\underline{\tilde{A}}^a\,\right)}}
={i\over{l^2}}\,\overline{\eta}_a\,,\label{Ashtek7}
\end{equation}
compare with (\ref{Hamilt34}), whereas the remaining ones $^\#\pi
^{u^0}$, $^\#\pi ^{\vartheta}_a $, $^\#\pi
^{^{\tilde{A}_{\bot}}}_a$ are all equal to zero. Proceeding as in
Subsection {\bf IV. D}, we build the total Hamiltonian 3--form
\begin{eqnarray}
{\cal H}&=&\,u^0\,\Bigl[\,{1\over{l^2}}\left(\, \vartheta
_a\wedge\underline{\tilde{F}}^a -\Lambda\,
\overline{\eta}\,\right)
-\tilde{A}_{\bot}^a\left(\underline{\tilde{D}}\,{}^\#\pi
^{^{\underline{\tilde{A}}}}_a +\epsilon _{ab}{}^c\,\vartheta
^b\wedge {}^\#\pi ^{\vartheta}_c\right)\,\Bigr]\nonumber \\
&&+\beta _{u^0}\,^\#\pi ^{u^0} +\beta _{\vartheta}^a\wedge{}^\#\pi
^{\vartheta}_a +\beta _{_{\tilde{A}_{\bot}}}^a\,^\#\pi
^{^{\tilde{A}_{\bot}}}_a +\beta
_{_{\underline{\tilde{A}}}}^a\wedge\Bigl(\,^\#\pi
^{^{\underline{\tilde{A}}}}_a
-{i\over{l^2}}\,\overline{\eta}_a\,\Bigr)\,.\label{Ashtek8}
\end{eqnarray}

Time evolution of any dynamical variable is calculable with the
help of the Poisson brackets (\ref{Hamilt17}) adapted to the
conjugate variables $u^0\,,{}^\#\pi ^{u^0}$; $\vartheta
^a\,,{}^\#\pi ^{\vartheta}_a$; $\tilde{A}_{\bot}^a\,,{}^\#\pi
^{^{\tilde{A}_{\bot}}}_a$; $\underline{\tilde{A}}^a\,,{}^\#\pi
^{^{\underline{\tilde{A}}}}_a$. Repeating the steps of Subsection
{\bf IV. D}, the stability conditions of the primary constraints
yield on the one hand the secondary constraints
\begin{eqnarray}
\varphi ^{^{(0)}}&:=&\,\vartheta _a \wedge\underline{\tilde{F}}^a
-\Lambda\,\overline{\eta}\,,\label{Ashtek9a}\\
\varphi ^{^{(1)}}_a&:=&\,\underline{\tilde{D}}\,{}^\#\pi
^{^{\underline{\tilde{A}}}}_a +\epsilon _{ab}{}^c\,\vartheta
^b\wedge {}^\#\pi ^{\vartheta}_c\,,\label{Ashtek9b}
\end{eqnarray}
and on the other hand the conditions on the Lagrange multipliers
\begin{equation}
i\,\overline{\eta}_{ab}\wedge\beta _{\vartheta}^b
-\underline{\tilde{D}}\,( u^0\,\vartheta _a )=
0\,,\label{Ashtek10}
\end{equation}
\begin{equation}
i\,\overline{\eta}_{ab}\wedge\beta^b_{_{\underline{\tilde{A}}}}
+u^0\,\Bigl(\, \underline{\tilde{F}}_a -\Lambda\,\overline{\eta}_a
\,\Bigr)= 0\,.\label{Ashtek11}
\end{equation}
From (\ref{Ashtek10}) follows
\begin{equation}
\beta_{\vartheta}^a =-i\,\left[\,e^a\rfloor
{}^\#\underline{\tilde{D}}\,(u^0\,\vartheta _b)\right]\vartheta ^b
+{i\over 2}\,\vartheta ^a\,{}^\#\left[\,\underline{\tilde{D}}\,(
u^0\,\vartheta _b )\wedge\vartheta ^b\right]\,,\label{Ashtek12}
\end{equation}
and from (\ref{Ashtek11}) with (\ref{Ashtek9a})
\begin{equation}
\beta_{_{\underline{\tilde{A}}}}^a
\approx\,i\,u^0\,\left[\,{}^\#\underline{\tilde{F}}^a
-e^a\rfloor\left(\,\vartheta
_b\wedge{}^\#\underline{\tilde{F}}^b\,\right)\,\right]\,.\label{Ashtek13}
\end{equation}
The constraint (\ref{Ashtek9b}) is stable. Instead,
(\ref{Ashtek9a}) requires the additional stability condition
\begin{equation}
\beta _{\vartheta}^a\wedge\left(\,\underline{\tilde{F}}_a
-\Lambda\overline{\eta}_a\,\right)+\vartheta
_a\wedge\underline{\tilde{D}}\,\beta _{_{\underline{\tilde{A}}}}^a
\approx 0\,,\label{Ashtek14}
\end{equation}
which, by making use of (\ref{Ashtek10}), (\ref{Ashtek11}) and
(\ref{Ashtek13}), transforms into
\begin{equation}
\underline{d}\left[\,(u^0)^2\,\left(\,\vartheta _a\wedge
{}^\#\underline{\tilde{F}}^a\right)\,\right]\approx
0\,.\label{Ashtek15}
\end{equation}
The stability of (\ref{Ashtek15}) requires
\begin{equation}
\vartheta _a\wedge{}^\#\underline{\tilde{F}}^a\approx
0\,,\label{Ashtek16}
\end{equation}
constituting a new constraint, replacing the --less restrictive--
previously found (\ref{Ashtek15}). Substitution of
(\ref{Ashtek16}) in (\ref{Ashtek13}) yields
\begin{equation} \beta_{_{\underline{\tilde{A}}}}^a
\approx\,i\,u^0\,{}^\#\underline{\tilde{F}}^a\,.\label{Ashtek17}
\end{equation}
Our search for constraints finishes by checking that
(\ref{Ashtek16}) is stable.

Let us at this point argue in favor of the equivalence between the
Ashtekar and the Hamiltonian PGT approach to gravity. The proof
requires first to support the strict Ashtekar character of the
present treatment; we achieve it by showing that, although
somewhat hidden by the exterior calculus notation, the already
obtained constraints satisfied by the complex variables
(\ref{Ashtek1}) coincide with the Ashtekar constraints. The second
step consists in demonstrating that the complex approach in terms
of (\ref{Ashtek1}) --that is, in terms of variables of the
Ashtekar type built from PGT quantities-- constitutes an
alternative formulation of the real Hamiltonian approach to PGT as
presented in {\bf IV. D, E}. Actually, we are going to show that,
by decomposing the dynamical equations of the Ashtekar kind into
their real and imaginary parts, they reproduce the Hamiltonian PGT
equations.

Our first task is to rewrite the constraints (\ref{Ashtek9a}),
(\ref{Ashtek9b}) and (\ref{Ashtek16}) in a language suitable to
reveal them as the well known Ashtekar constraints. (When
comparing the following results with the standard equations, for
instance (4) and (6) of \cite{Ashtekar:1996qw}, the reader must
have in mind the interchanged role of the latin letters in
Ashtekar's notation for indices as compared with ours, being in
our case those of the beginning of the alphabet reserved for
internal $SO(3)$ indices, see Appendix {\bf C}, while those of the
middle of the alphabet are assigned by us to the general
coordinate indices of the underlying four--dimensional manifold.)
Let us begin with the constraint (\ref{Ashtek9b}), transforming
with the help of the primary constraints ${}^\#\pi ^{\vartheta}_a$
and (\ref{Ashtek7}) into $\varphi ^{^{(1)}}_a\approx {i\over
{l^2}}\,\underline{\tilde{D}}\,\overline{\eta}_a$. Its
3--dimensional Hodge dual manifests itself as the Gauss law
\begin{equation}
{\cal G}_a
:={}^\#\left(\underline{\tilde{D}}\,\overline{\eta}_a\,\right)
=\,{1\over e}\,\tilde{D}{}_i \left( e\,
e_a{}^i\right)\approx\,0\,,\label{Ashtek22}
\end{equation}
with $e$ as the determinant built from the components of the triad
$\vartheta ^a=dx^i e_i{}^a$. Similarly, from (\ref{Ashtek16}) we
get
\begin{equation}
{}^\#\Bigl(\,\vartheta
_a\wedge\,{}^\#\underline{\tilde{F}}^a\,\Bigr)=\vartheta ^b
\tilde{F}_{ba}^a =dx^i\tilde{F}_{ia}^a\approx 0 \Rightarrow \qquad
{\cal V}_i :=
e_a{}^j\,\tilde{F}_{ij}{}^a\approx\,0\,,\label{Ashtek23}
\end{equation}
where one recognizes Ashtekar's vector constraint. Finally, the
Hodge dual of (\ref{Ashtek9a}) becomes the ordinary scalar
constraint, namely
\begin{equation}
{\cal S}:={}^\#\Bigl(\vartheta _a\wedge\underline{\tilde{F}}^a
-\Lambda\,\overline{\eta}\,\Bigr) =\,{1\over 2}\, \epsilon
_a{}^{bc}\,\tilde{F}^a_{bc} -\Lambda =\,{1\over 2}\, \epsilon
_a{}^{bc}\, e_b{}^i\, e_c{}^j\,\tilde{F}^a_{ij} -\Lambda
\approx\,0\,.\label{Ashtek24}
\end{equation}
In view of (\ref{Ashtek22})--(\ref{Ashtek24}), the full
identification of (\ref{Ashtek1}) with the Ashtekar variables will
be complete once the spin connection, and thus (\ref{Ashtek1})
itself, becomes entirely determined by the coframe as a
consequence of the vanishing of torsion, as will be shown below.

On the other hand, the announced proof of the exact coincidence
between the present results and the PGT ones in {\bf IV. D, E}
requires to reproduce here the dynamical equations
(\ref{Hamilt60}) and (\ref{Hamilt61}) together with the zero
torsion conditions (\ref{Hamilt57a}), (\ref{Hamilt57b}). We
proceed as follows. From (\ref{Ashtek8}) we calculate the
evolution equations for $\underline{\tilde{A}}^a\,$ to be
\begin{equation}
u^0\,{\it{l}}_{e_{_0}}\,\underline{\tilde{A}}^a
=\beta_{_{\underline{\tilde{A}}}}^a
+\underline{\tilde{D}}\,(u^0\tilde{A}_{\bot})\,.\label{Ashtek18}
\end{equation}
Taking the value (\ref{Ashtek17}) into account with the first
definition in (\ref{Ashtek6}), from (\ref{Ashtek18}) we get
\begin{equation}
\tilde{F}_{\bot}^a
=\,i\,{}^\#\underline{\tilde{F}}^a\,.\label{Ashtek19}
\end{equation}
Equation (\ref{Ashtek19}) can be rewritten in 4--dimensional
notation by recalling
\begin{equation}
\tilde{F}^a =\hat{\vartheta}^0\wedge\tilde{F}_{\bot}^a
+\underline{\tilde{F}}^a\,\,,\qquad {}^*\tilde{F}^a
=\hat{\vartheta}^0\wedge {}^\#\underline{\tilde{F}}^a
-{}^\#\tilde{F}_{\bot}^a\,,\label{fieldstrdecomp}
\end{equation}
according to (\ref{Hamilt4}) and (\ref{Hamilt6}) respectively. So,
in four dimensions, (\ref{Ashtek19}) transforms into
\begin{equation}
{}^*\tilde{F}^a =-i\,\tilde{F}^a\,,\label{Ashtek20}
\end{equation}
establishing a simple relation between the field strength and its
Hodge dual. (By the way, notice that (\ref{Ashtek20}) guarantees
the automatic fulfilment of the gauge theoretical equation
$\tilde{D}\,{}^*\tilde{F}^a =-i\,\tilde{D}\tilde{F}^a\equiv 0$.)
Furthermore, (\ref{Ashtek9a}) and (\ref{Ashtek19}) with
(\ref{Ashtek16}) yield
\begin{equation}
\hat{\vartheta}_a\wedge {\tilde{F}}^a -\Lambda\,\hat{\eta}_{_0}
=\,0\,.\label{Ashtek21}
\end{equation}

Let us show that (\ref{Ashtek20}) and (\ref{Ashtek21}) constitute
an alternative way to display the previously found PGT equations
(\ref{Hamilt60}) and (\ref{Hamilt61}) respectively. Indeed, taking
(\ref{Ashtek2b}) into account one checks that (\ref{Ashtek20}) is
a shorthand for $\left( D\,X^a -{}^*{\cal{R}}^a\,\right)
-i\,{}^*\left( D\,X^a -{}^*{\cal{R}}^a\,\right)\approx\,0\,$,
doubly reproducing (\ref{Hamilt60}), while (\ref{Ashtek21}) can be
rewritten as $\left( \hat{\vartheta}_a\wedge {\cal{R}}^a
-\Lambda\,\hat{\eta}_{_0}\,\right) -i\,\hat{\vartheta}_a\wedge
DX^a \approx\,0\,$. The imaginary contribution, reexpressed in
terms of the torsion components (\ref{App2.21a}),
(\ref{App2.21b}), reads $\hat{\vartheta}_a\wedge DX^a\equiv
-d\hat{T}^0 +\hat{T}_a\wedge X^a\,$, so that provided the torsion
vanishes, (\ref{Ashtek21}) reproduces (\ref{Hamilt61}). If this is
the case, the dynamical equations derived from both approaches
coincide.

The result of zero torsion follows in fact on the one hand from
(\ref{Ashtek9b}), reduced in view of the primary constraints
${}^\#\pi ^{\vartheta}_a\approx 0$ and (\ref{Ashtek7}) to $\varphi
^{^{(1)}}_a\approx {i\over
{l^2}}\,\underline{\tilde{D}}\,\overline{\eta}_a = {i\over
{l^2}}\,\overline{\eta}_{ab}\wedge\underline{\tilde{D}}\,\vartheta
^b\approx 0$, and on the other hand from (\ref{Ashtek10}) with the
value of $\beta _{\vartheta}^b$ given by the evolution equation
for the triad, namely $\beta _{\vartheta}^b = u^0\,\tilde{{\cal
\L\/}}_{\hat{e}_{_0}}\vartheta ^b\,$. Combining both results into
a single four--dimensional expression in order to facilitate
calculations, we get
\begin{equation}
{1\over{u^0}}\,\vartheta ^0\wedge\left[\,
i\,\overline{\eta}_{ab}\wedge u^0\,\tilde{{\cal
\L\/}}_{\hat{e}_{_0}}\vartheta ^b -\underline{\tilde{D}}\,(
u^0\,\vartheta _a )\,\right]
-i\,\overline{\eta}_{ab}\wedge\underline{\tilde{D}}\,\vartheta ^b
=\tilde{D}\left(\,\vartheta ^0\wedge\vartheta ^a +i\,\eta
^{0a}\,\right)\approx 0\,,\label{suplem1}
\end{equation}
with $\tilde{D}$ as the $SO(3)$ covariant derivative built with
the complex connection (\ref{Ashtek1}). Taking into account the
expressions (\ref{App2.21a}) and (\ref{App2.21b}) for torsion, we
find
\begin{equation}
0\approx \tilde{D}\left(\,\vartheta ^0\wedge\vartheta ^a +i\,\eta
^{0a}\,\right) = T^0\wedge\vartheta ^a -\vartheta ^0\wedge T^a
+i\,\eta ^{0a}{}_b\wedge T^b\,,\label{suplem2}
\end{equation}
whose unique solution is the vanishing of the whole torsion
$T^\alpha\,$.

Observe that zero torsion allows to simplify (\ref{Ashtek12})
enormously provided one expresses it as
\begin{equation}
\beta_{\vartheta}^a =-u^0\,\underline{X}^a -i\,e^a\rfloor
{}^\#\underline{d}\,u^0\,,\label{suplem4}
\end{equation}
that is, in terms of the real and imaginary parts of
(\ref{Ashtek1}) separately, rather than in terms of the whole
complex connection (\ref{Ashtek1}). Compare (\ref{suplem4}) with
(\ref{Hamilt39}).

A more relevant consequence of $T^\alpha\approx 0\,$ is that
(\ref{Ashtek1}) becomes expressible in terms of the torsion free
connection $\hat{\Gamma}^{\{\}}_{\mu\nu}$ of the form displayed in
(\ref{Christoffel}) as
\begin{equation}
\tilde{A}^a =-{i\over 2}\,\left[ e^\mu\rfloor e^\nu\rfloor
\left(\,\vartheta ^0\wedge\vartheta ^a +i\,\eta
^{0a}\,\right)\,\right]\,\hat{\Gamma}^{\{\}}_{\mu\nu}\,\,\, =\,
{1\over 2}\,\epsilon ^a{}_{bc}\hat{\Gamma}_{\{\}}^{bc}
+i\,\hat{\Gamma}_{\{\}}^{0a}\,,\label{suplem3}
\end{equation}
see (\ref{App2.16b}), (\ref{App2.16c}). This completes the
correspondence between Hamiltonian PGT built exclusively in terms
of real quantities as developed before, and the present
Hamiltonian treatment in terms of complex Ashtekar variables, the
latter satisfying the Ashtekar constraints
(\ref{Ashtek22})--(\ref{Ashtek24}) and being built from coframes
as shown in (\ref{suplem3}). We claim that the identification of
the Ashtekar complex connection as the combination (\ref{Ashtek1})
of the PGT real fields (\ref{App2.16b}) and (\ref{App2.16c})
allows to regard both Hamiltonian approaches to gravity
--Ashtekar's and PGT-- as alternative reformulations of each
other.

\section{Metric-affine gravity}

Finally, let us briefly illustrate the nonlinear techniques when
applied to a spacetime group other than the Poincar\'e group. We
consider in particular the affine group giving rise to
metric--affine gravity (MAG) \cite{Hehl:1995ue}
\cite{Julve:1994bh} \cite{Julve:1995yf} \cite{Lopez-Pinto:1995qb}
\cite{Tresguerres:2000qn}, which constitutes an open and active
research field --see \cite{Hehl:1995ue} \cite{Hehl:1999sb} and
references therein-- proposed as an alternative to more usual
descriptions of gravity. In the various nonlinear approaches to
PGT studied in previous sections, we remarked the interpretation
of tetrads as gauge--theoretical quantities, specifically as
nonlinear translative connections. This result remaining valid in
the context of the MAG theory to be presented here, we are going
to pay further attention to the origin of the degrees of freedom
of the MAG--metric, which also turn out to be of
gauge--theoretical nature as Goldstone fields. To make this point
apparent, we consider two different nonlinear approaches to the
affine group $G=A(4\,,R)$, corresponding to the choices of the
auxiliary subgroup either as the general linear group
$H_1=GL(4\,,R)$ or as the homogeneous Lorentz group
$H_2=SO(3\,,1)$, and then we relate them to each other. The reason
for doing so is that by simply applying the standard nonlinear
gauge procedure to the affine group with $H_1=GL(4\,,R)$, no
metric tensor becomes manifest. For the latter to be deduced, the
formalism obtained for $H_1=GL(4\,,R)$ has to be compared to the
one derived for $H_2=SO(3\,,1)$, as will be shown immediately.

Let us start with the nonlinear realization of $G=A(4\,,R)$ with
$H_1=GL(4\,,R)$. Proceeding as usual, we replace in the simplified
form (\ref{simplif}) of (\ref{nonlintrans2}) the suitable $G$
elements
\begin{equation}
g=\,e^{i\,\epsilon ^{\alpha} P_\alpha} e^{i\,\omega _\alpha
{}^\beta \Lambda   ^\alpha {}_\beta} \approx I+i\,\epsilon
^{\alpha} P_\alpha +i\,\omega _\alpha {}^\beta \Lambda ^\alpha
{}_\beta\,,\label{param}
\end{equation}
with infinitesimal transformation parameters $\epsilon ^\alpha$
and $\omega _\alpha {}^\beta$, as much as the $H_1$ elements
\begin{equation}
h:=e^{i\,v _\alpha {}^\beta \Lambda   ^\alpha {}_\beta} \approx
I+i\,v _\alpha {}^\beta \Lambda   ^\alpha {}_\beta \,,\label{infh}
\end{equation}
with the also infinitesimal group parameter $v _\alpha {}^\beta$,
and
\begin{equation}
\tilde{b}=e^{-i\,\xi ^\alpha P_\alpha}\,,\qquad
\tilde{b}\,'\,=e^{-i\,(\,\xi ^\alpha +\delta\xi ^\alpha\,)
P_\alpha} \,,\label{zerosec}
\end{equation}
where the finite translational parameters $\xi ^\alpha$ label the
cosets $\tilde{b}\in G/H_1$. The tildes in (\ref{zerosec}) are
introduced to distinguish the tilded $\tilde{b}$'s from the
untilded $b$ in (\ref{fact}) below. Using the Hausdorff--Campbell
formula (\ref{HC1}) with the commutation relations of the affine
group
\begin{eqnarray}
\left[P_{\alpha}\, , P_{\beta}\right] &=& 0,\nonumber \\
\left[\Lambda ^{\alpha}{}_{\beta}\, ,  P_{\gamma}\right] &=&
\delta^{\alpha}_{\gamma}\, P_{\beta}\, , \nonumber\\
 \left[\Lambda ^{\alpha}{}_{\beta},  \Lambda ^{\gamma}{}_{\delta}\right] &=&
\delta^{\alpha}_{\delta}\, \Lambda ^{\gamma}{}_{\beta} -
\delta^{\gamma}_{\beta}\, \Lambda
^{\alpha}{}_{\delta}\,,\label{affcomrel}
\end{eqnarray}
we find the value of $v _\alpha{}^\beta$ in (\ref{infh}) and the
variation of the coset parameters $\xi ^{\alpha}$ in
(\ref{zerosec}) to be respectively
\begin{equation}
v _\alpha {}^\beta =\,\omega _\alpha {}^\beta\,,\qquad \delta\xi
^{\alpha }=-\xi ^\beta\omega _\beta {}^\alpha -\epsilon
^\alpha\,,\label{affcov}
\end{equation}
compare with the analogous PGT results (\ref{paramvars}). The
nonlinear connection (\ref{compar4}) is built in terms of the
linear affine connection
\begin{equation}
A_{_M}:=-i\,{\buildrel ({\rm T})\over{\Gamma ^\alpha}} P_\alpha
-i\,{\buildrel ({\rm GL})\over{\Gamma _\alpha {}^\beta }} \Lambda
^\alpha {}_\beta \,,\label{lincon}
\end{equation}
whose components, the translational and the $GL(4\,,R)$
connection, transform respectively as
\begin{equation}
\delta {\buildrel ({\rm T})\over{\Gamma ^{\alpha}}} = - {\buildrel
({\rm T})\over{\Gamma ^{\beta}}}\omega _\beta {}^\alpha
+{\buildrel ({\rm GL})\over{D}}\epsilon ^\alpha\,,\qquad \delta
{\buildrel ({\rm GL})\over{\Gamma _\alpha {}^\beta }}=\,
{\buildrel ({\rm GL})\over{D}}\omega _\alpha {}^\beta
\,.\label{lintr}
\end{equation}
Replacing (\ref{lincon}) in (\ref{compar4}) we get
\begin{equation}
\tilde{\Gamma}_{_M}:=\,\tilde{b}^{-1}\bigl(\,d\,+A_{_M}\bigr)\tilde{b}
=-i\,\tilde{\vartheta}^\alpha P_\alpha -i\,\tilde{\Gamma}_\alpha
{}^\beta \Lambda ^\alpha{}_\beta\,,\label{affnlconn}
\end{equation}
where
\begin{equation}
\tilde{\vartheta }^\alpha :={\buildrel ({\rm GL})\over D}\xi
^\alpha +{\buildrel ({\rm T})\over{\Gamma ^\alpha}}\,,\qquad
\tilde{\Gamma }_\alpha {}^\beta ={\buildrel ({\rm GL})
\over{\Gamma _\alpha {}^\beta }}\,.\label{deftetr}
\end{equation}
As in the case of (\ref{zerosec}), we denote these objects with a
tilde for later convenience. Applying (\ref{varnlcon}), it is
trivial to find
\begin{equation}
\delta \tilde{\vartheta }^{\alpha} =- \tilde{\vartheta
}^{\beta}\omega _\beta {}^\alpha\,,\qquad \delta \tilde{\Gamma
}_\alpha {}^\beta =\, \tilde{D}\omega _\alpha
{}^\beta\,,\label{modtransfs}
\end{equation}
showing that the coframe $\tilde{\vartheta }^\alpha$ in
(\ref{deftetr}) transforms as a $GL(4\,,R)$ covector, in contrast
to the linear translational connection, see (\ref{lintr}), while
$\tilde{\Gamma}_\alpha {}^\beta$ remains unchanged as a
$GL(4\,,R)$ connection.

So far, no metric tensor is derived from the gauging of the affine
group. In order to deductively obtain a metric as a
gauge--theoretical quantity, we have to consider a second
nonlinear realization of $G=A(4\,,R)$ with auxiliary subgroup
$H_2=SO(3\,,1)$. Being the homogeneous Lorentz group a
(pseudo-)orthogonal group, it is equipped with a Cartan-Killing
metric, namely the invariant Minkowski metric $o_{\alpha\beta }$.
When taken as the auxiliary subgroup of the nonlinear realization,
the Lorentz group induces an automatic metrization of the theory.
Certainly, as long as we only attend to the realization with
$H_2=SO(3\,,1)$, the metric can just be a constant, the Lorentz
invariance $\delta o_{\alpha\beta }=\,0$ still holding under gauge
transformations of the whole affine group since, as a general
feature of the nonlinear procedure, the total group $G$ acts
formally as its subgroup $H_2$, see (\ref{varmatt}). Nevertheless,
we are going to show how to establish the correspondence to the
realization with $H_1=GL(4\,,R)$ studied above, in such a way
that, by means of redefinitions isomorphic to gauge
transformations, ten Goldstone--like degrees of freedom may be
either rearranged in the gauge potentials or displayed as a
variable metric tensor, depending on the nonlinear realization we
consider, either with $H_2=SO(3\,,1)$ or with $H_1=GL(4\,,R)$, see
(\ref{corrb})--(\ref{lineelement}) below.

To study the case with $H_2=SO(3\,,1)$, we start by splitting the
generators of the general linear group into the sum of symmetric
plus antisymmetric (Lorentz) parts as $\Lambda ^\alpha {}_\beta
=\,S^\alpha {}_\beta + L^\alpha {}_\beta$. Then we apply the
general formula (\ref{simplif}) with the particular
parametrization
\begin{equation}
g=\,e^{i\,\epsilon ^{\alpha} P_\alpha} e^{i\,\alpha
_\alpha{}^\beta S^\alpha{}_\beta} e^{i\,\beta _\alpha{}^\beta
L^\alpha{}_\beta}\quad\,,\quad b:=e^{-i\,\xi ^\alpha
P_\alpha}e^{i\,h_\alpha{}^\beta S^\alpha{}_\beta} \quad\,,\quad
h:=e^{i\,u _\alpha{}^\beta L^\alpha{}_\beta}\quad\,,\label{fact}
\end{equation}
where the infinitesimal transformation parameters of the affine
group are the translative ones $\epsilon ^{\alpha }$ and the
general linear parameters in (\ref{param}), decomposed into
symmetric plus antisymmetric contributions as $\omega
_\alpha{}^\beta =\alpha _\alpha{}^\beta +\beta _\alpha{}^\beta$,
while the infinitesimal nonlinear parameters $u _\alpha{}^\beta$
correspond to the Lorentz subgroup $H_2$. From (\ref{simplif}) we
find the variations ${\xi\,'}^\alpha =\,\xi ^{\alpha} +\delta\xi
^{\alpha}$ and $h'_\alpha{}^\beta =\,h_\alpha{}^\beta +\delta
h_\alpha{}^\beta$ of the coset parameters of $b$ in (\ref{fact})
to be respectively
\begin{equation}
\delta\xi ^{\alpha }=-\xi ^\beta\left(\alpha _\beta {}^\alpha
+\beta _\beta {}^\alpha \right)-\epsilon ^\alpha\,,\qquad \delta
r_\alpha{}^\beta =\,\left(\,\alpha _\alpha{}^\gamma +\beta
_\alpha{}^\gamma\,\right)\, r_\gamma{}^\beta - u_\gamma{}^\beta\,
r_\alpha{}^\gamma\,\,,\label{affcovb}
\end{equation}
where we made use of the definition
\begin{equation}
r_\alpha{}^\beta :=\,\left(\,e^h\right) _\alpha{}^\beta :=\delta
_\alpha{}^\beta +h_\alpha{}^\beta +{1\over{2!}}\,h_\alpha{}^\gamma
h_\gamma{}^\beta +\cdots\label{rtrans}
\end{equation}
It is easy to check that, contrary to $\mu _\alpha{}^\beta$ in
(\ref{paramvars}), the nonlinear Lorentz parameters
$u_\alpha{}^\beta$ relevant for nonlinear transformations differ
from the linear ones $\beta _\alpha{}^\beta$. But we do not need
to know their explicit form, which can be calculated from the
vanishing of the antisymmetric part of the second equation in
(\ref{affcovb}), see \cite{Lopez-Pinto:1995qb}
\cite{Tresguerres:2000qn}.

The nonlinear connection corresponding to the choice
$H_2=SO(3\,,1)$ is obtained by replacing (\ref{lincon}) into
(\ref{compar4}) taking into account the decomposition $\Lambda
^\alpha {}_\beta =S^\alpha {}_\beta + L^\alpha {}_\beta$, with $b$
as given in (\ref{fact}). We get
\begin{equation}
\Gamma _{_M}:=\,b^{-1}\bigl(\,d\,+A_{_M}\bigr)b =-i\,\vartheta
^\alpha P_\alpha -i\,\Gamma _\alpha {}^\beta \Bigl( S^\alpha
{}_\beta +L^\alpha {}_\beta\Bigr)\,,\label{maglincon}
\end{equation}
with
\begin{equation}
\vartheta ^\alpha :=\Bigl({\buildrel ({\rm GL})\over{D}}\xi ^\beta
+{\buildrel ({\rm T})\over{\Gamma ^\beta}}\,\Bigr) r_\beta
{}^\alpha \,,\qquad \Gamma _\alpha {}^\beta
:=\,\left(\,r^{-1}\right)_\alpha {}^\gamma \Bigl[\,{\buildrel
({\rm GL})\over{\Gamma _\gamma {}^\lambda }}\, r_\lambda {}^\beta
-d\,r_\gamma {}^\beta\,\Bigr]\,.\label{noncon}
\end{equation}
The coframe $\vartheta ^\alpha $ in (\ref{noncon}) transforms as a
Lorentz covector, that is
\begin{equation}
\delta\vartheta ^{\alpha } =-\vartheta ^\beta\,u_\beta {}^\alpha
\,,\label{Locov}
\end{equation}
with $u_\beta {}^\alpha$ as the nonlinear Lorentz parameters,
whereas the linear connection in (\ref{noncon}), taken as a whole,
behaves as a Lorentz connection
\begin{equation}
\delta \Gamma _\alpha {}^\beta =\,D\,u_\alpha {}^\beta
\,.\label{Locon}
\end{equation}
Observe however that, as read out from the r.h.s. of
(\ref{maglincon}), the decomposition into two sectors of the Lie
algebra of $GL(4\,,R)$ gives rise to a splitting of the linear
connection into the sum of a symmetric plus an antisymmetric part.
Only the latter, with values on the Lorentz algebra, behaves as a
true Lorentz connection, while the symmetric part (that is, the
nonmetricity $Q_{\alpha\beta}:=2\,\Gamma _{(\alpha\beta )}$) is a
Lorentz tensor, varying as
\begin{equation}
\delta Q_{\alpha\beta}=2\,u_{(\alpha}{}^\gamma Q_{\beta
)\gamma}\,.\label{trnm}
\end{equation}

Having completed the nonlinear realization of the affine group
with the auxiliary subgroup $H_2$, we are ready to establish the
correspondence between it and the one with $H_1$. The affine
objects of the approach with $H_1=GL(4\,,R)$ are displayed in
(\ref{deftetr}), distinguished by tildes, while those of the
$H_2=SO(3\,,1)$ case are written without tildes in (\ref{noncon}).
By comparing (\ref{deftetr}) and (\ref{noncon}) to each other, we
find out that the relation between both kinds of quantities is
isomorphic to a finite gauge transformation expressible as
\begin{equation}
\tilde{\vartheta }^\alpha = {\buildrel ({\rm GL})\over{D}}\xi
^\alpha + {\buildrel ({\rm T})\over{\Gamma ^\alpha}} =\,\vartheta
^\beta \left(\,r^{-1}\right) _\beta {}^\alpha \,,\label{corrb}
\end{equation}
and
\begin{equation}
\tilde{\Gamma }_\alpha {}^\beta = \,{\buildrel ({\rm
GL})\over{\Gamma _\alpha {}^\beta }}
=\,r_\alpha{}^\gamma\Bigl[\,\Gamma _\gamma {}^\lambda\left(
r^{-1}\right)_\lambda {}^\beta -d\,\left( r^{-1}\right)_\gamma
{}^\beta\,\Bigr]\,,\label{corra}
\end{equation}
with the main difference that the matrix $r_\alpha{}^\beta$ as
given by (\ref{rtrans}) is not a gauge transformation matrix, but
consists of coset fields varying as shown in (\ref{affcovb}). It
is precisely this peculiar transformation property of
$r_\alpha{}^\beta$, involving both the linear ($\omega
_\alpha{}^\beta =\alpha _\alpha{}^\beta +\beta _\alpha{}^\beta$)
as much as the nonlinear group parameters ($u _\alpha{}^\beta$),
that is responsible for the difference between the gauge
transformations --(\ref{modtransfs}) {\it{versus}} (\ref{Locov}),
(\ref{Locon})-- of the objects with and without tilde
respectively, related to each other by $r_\alpha{}^\beta$ as
displayed in (\ref{corrb}) and (\ref{corra}).

In analogy to the latter equations, a correspondence can be
established between the Minkowski metric $o_{\alpha\beta}$,
existing in the $H_2=SO(3\,,1)$ approach as a natural invariant,
and a correlated MAG-metric tensor $\tilde{g}_{\alpha\beta }$
defined in the context of the approach with $H_1=GL(4\,,R)$ as
\begin{equation}
\tilde{g}_{\alpha\beta }:=\,r_\alpha {}^\mu r_\beta {}^\nu
o_{\mu\nu }\,.\label{metr}
\end{equation}
The MAG-metric tensor (\ref{metr}) plays the role of a Goldstone
field \cite{Borisov74} \cite{Tresguerres:2000qn}. Actually, the
ten degrees of freedom associated to it drop out by inverting the
"gauge transformation" (\ref{metr}), together with (\ref{corrb})
and (\ref{corra}). In other words, it is possible to absorb the
metric variables into redefined gauge potentials by using the
nonlinear realization with the Lorentz group $H_2=SO(3\,,1)$ as
the auxiliary subgroup instead of $H_1=GL(4\,,R)$. Accordingly,
affine invariants can be alternatively displayed in terms of
explicit general linear quantities (with tildes), as in the
standard formulation of MAG \cite{Hehl:1995ue}, or in terms of
explicit Lorentz objects (without tildes) with the metric fixed to
be Minkowskian. For instance, the line element can be doubly
expressed as
\begin{equation}
ds^2 =\,\tilde{g}_{\alpha\beta}\tilde{\vartheta }^\alpha
\otimes\tilde{\vartheta }^\beta =\,o_{\alpha\beta }\vartheta
^\alpha\otimes\vartheta ^\beta\,.\label{lineelement}
\end{equation}
In accordance with the Goldstone--like nature of the MAG--metric,
the field equations obtained by varying affine invariant actions
with respect to $\tilde{g}_{\alpha\beta}$ are known to be
redundant \cite{Hehl:1995ue}. However, we wont enter the study of
details concerning MAG--dynamics. The interested reader is
referred to the literature, where quite general actions were
studied, involving quadratic curvature, torsion and nonmetricity
terms, for which a number of exact solutions were found
\cite{Tresguerres:1995js} \cite{Tresguerres:1995un}
\cite{Vlachynsky:1996zh} \cite{Obukhov:1997ka}
\cite{Garcia:1998jw} \cite{Hehl:1998nx} \cite{Socorro:1998fm}
\cite{Puetzfeld:2001hk}. Discussions on the problem of the
inclusion of matter sources in the MAG scheme can also be found
for instance in references \cite{Obukhov:1993pt}
\cite{Obukhov:1996mg} \cite{Babourova:1995fv}
\cite{Babourova:1996xe} \cite{Ne'eman:1997iz} dealing with
phenomenological matter, and in \cite{Hehl:1995ue}
\cite{Lopez-Pinto:1996pu} \cite{Ne'eman:1997uc} concerning
fundamental matter.

\section{Conclusions}

We presented a number of applications of NLR's to the foundation
of different gravitational gauge theories in order to illustrate
the variety of fields in which the method reveals to be useful,
providing underlying mathematical unity and simplicity. In
particular, it is worth to recall once more that the Hamiltonian
approach developed in Section {\bf IV} in terms of PGT connection
variables revealed to be dynamically equivalent to a theory of the
Ashtekar type. (So that, conversely, the latter can be regarded as
a reformulation of the Hamiltonian Poincar\'e gauge theory built
from the Einstein--Cartan action.)

As a general result derived from the different examples studied by
us, we want to remark that thanks the NLR's the description of
interactions is achieved exclusively in terms of connections, in
accordance with the general gauge--theoretical program. Neither
the coframes nor the MAG--metric are to be regarded as separate
gravitational potentials of specific nature, but rather as
ordinary Yang-Mills objects. Indeed, the coframes are interpreted
as a kind of gauge potentials, namely as nonlinear translative
connections, while the metric of MAG is found to be a Goldstone
field playing no fundamental physical role, since its degrees of
freedom can be transferred to redefined gauge potentials. In the
limit of vanishing Poincar\'e connections (corresponding to zero
gravitational forces), the tetrads (\ref{tetrad}) reduce to the
special relativistic ones $\vartheta ^\alpha =\,d\xi ^\alpha $,
the fields $\xi ^\alpha$ playing the role of ordinary coordinates
--as read out from their transformations (\ref{paramvars})--, so
that the Minkowski space of Special relativity can be seen as the
{\em residual} structure left by the dynamical theory of spacetime
when gravitational interactions are switched off.

Matter sources in the context of NLR's were exemplified by Dirac
fields in PGT, whose coupling to translations gives rise to a
background fermion mass contribution. Instead, the inclusion of
fundamental matter in the context of nonlinear metric--affine
gravity remains only partially explored. A further natural
extension of the nonlinear method not yet developed consists in
its application to mechanisms of spontaneous symmetry breaking
--from $G$ to a residual symmetry $H\subset G$-- in the case of
external as much as of internal groups. Let us also hope that,
although restricted for the time being to classical aspects of
gravity, the nonlinear framework can become an useful tool to deal
with quantum aspects of gravitational gauge theories.


\begin{acknowledgments}
The authors want to thank F.W. Hehl and E.W. Mielke for permanent
interest and encouragement along the years.
\end{acknowledgments}


\appendix
\section{Eta basis}

The objects constituting the eta basis defined in the present
Appendix, built as the Hodge duals of exterior products of tetrads
\cite{Hehl:1995ue}, are convenient to simplify the notation when
dealing with differential forms. In terms of (\ref{tetrad}) we
define the Levi-Civita object (that is, the $0$-form element of
the eta basis) as
\begin{equation}
\eta ^{\alpha\beta\gamma\delta}:=\,^*(\vartheta
^\alpha\wedge\vartheta ^\beta\wedge\vartheta
^\gamma\wedge\vartheta ^ \delta\,)\,,\label{levicivita}
\end{equation}
and with the help of it the $1$-form element
\begin{equation}
\eta ^{\alpha\beta\gamma}:=\,^*(\vartheta ^\alpha\wedge\vartheta
^\beta\wedge\vartheta ^\gamma\,)=\,\eta
^{\alpha\beta\gamma}{}_\delta\,\vartheta ^
\delta\,,\label{antisym1form}
\end{equation}
the eta-basis $2$-form element
\begin{equation}
\eta ^{\alpha\beta}:=\,^*(\vartheta ^\alpha\wedge\vartheta
^\beta\,)={1\over{2!}}\,\eta ^{\alpha\beta}{}_{\gamma\delta}
\,\vartheta ^\gamma\wedge\vartheta ^ \delta\,,\label{antisym2form}
\end{equation}
the $3$-form element (dual of the tetrad)
\begin{equation}
\eta ^\alpha :=\,^*\vartheta ^\alpha ={1\over{3!}}\,\eta
^\alpha{}_{\beta\gamma\delta} \,\vartheta ^\beta\wedge\vartheta
^\gamma\wedge\vartheta ^ \delta\,,\label{antisym3form}
\end{equation}
and the $4$-form of the eta-basis, or four--dimensional volume
element
\begin{equation}
\eta :=\,^*1 ={1\over{4!}}\,\eta
_{\alpha\beta\gamma\delta}\,\vartheta ^\alpha\wedge\vartheta
^\beta\wedge\vartheta ^\gamma\wedge\vartheta ^
\delta\,.\label{eta4form}
\end{equation}

The exterior product of the coframe $\vartheta ^\mu$ with the
elements (\ref{levicivita})--(\ref{antisym3form}) of the eta basis
yields respectively
\begin{eqnarray}
\vartheta ^\mu\wedge\eta _{\alpha\beta\gamma\delta} &=&-\delta
^\mu _\alpha\,\eta _{\beta\gamma\delta} +\delta ^\mu _\delta\,\eta
_{\alpha\beta\gamma} -\delta ^\mu _\gamma\,\eta
_{\delta\alpha\beta} +\delta ^\mu _\beta\,\eta
_{\gamma\delta\alpha}\,,\label{product1}\\
\vartheta ^\mu\wedge\eta _{\alpha\beta\gamma} &=&\delta ^\mu
_\alpha\,\eta _{\beta\gamma} +\delta ^\mu _\gamma\,\eta
_{\alpha\beta} +\delta ^\mu _\beta\,\eta _{\gamma\alpha}
\,,\label{product2}\\
\vartheta ^\mu\wedge\eta _{\alpha\beta} &=&-\delta ^\mu
_\alpha\,\eta _{\beta} +\delta ^\mu _\beta\,\eta _{\alpha}
\,,\label{product3}\\
\vartheta ^\mu\wedge\eta _{\alpha} &=&\delta
^\mu _\alpha\,\eta\,.\label{product4}
\end{eqnarray}
For an arbitrary p--form $\alpha$ on four--dimensional space with
Lorentzian signature, the double application of the Hodge dual
operator reproduces $\alpha$ itself up to the sign as $^{**}\alpha
=(-1)^{p(4-p)+1}\,\alpha$. A further relation involving Hodge
duality reads $^*(\alpha\wedge\vartheta
_\mu\,)=e_\mu\rfloor\,^*\alpha$, while for differential forms
$\alpha$, $\beta$ of the same degree p, equation
$^*\alpha\wedge\beta =\,^*\beta\wedge\alpha$ holds. The eta basis
(\ref{levicivita})--(\ref{eta4form}) and the algebraic relations
of the present Appendix are extensively used thorough the whole
work.

\section{\bf Hausdorff--Campbell formulas}

In order to make the present exposition as self contained as
possible, we give the well known formulas
\begin{equation}
e^{-A} B e^A =\,B-\left[ A\,,B\,\right] +{1\over{2!}}\left[
A\,,\left[ A\,,B\,\right]\,\right] -{1\over{3!}}\left[ A\,,\left[
A\,,\left[ A\,,B\,\right] \,\right]\,\right] +...\label{HC1}
\end{equation}
\begin{equation}
e^{-A}de^A =\,dA-{1\over{2!}}\left[ A\,,dA\,\right]
+{1\over{3!}}\left[ A\,,\left[ A\,,dA\,\right] \,\right]
-...\label{HC2}
\end{equation}
\begin{equation}
e^{A+\delta A}=\,e^A +\delta e^A +O\left(\left(\delta A\,\right)
^2\,\right) =\,e^A\left( 1+e^{-A}\delta e^A\,\right)\,,\label{HC3}
\end{equation}
useful for checking calculations.

\section{\bf Nonlinear realization of the Poincar\'e group with
$SO(3)$ as auxiliary subgroup}

By decomposing the Lorentz generators $L_{\alpha\beta}$ into
boosts $K_a$ and space rotations $S_a\,$, defined respectively as
\begin{equation}
K_a :=\,2\,L_{a0}\quad\,,\quad S_a :=-\epsilon _a{}^{bc}
L_{bc}\qquad\qquad (a=\,1\,,2\,,3)\,,\label{App2.1}
\end{equation}
the commutation relations (\ref{comrelpoinc}) transform into
\begin{eqnarray}
\left[ S_a\,,S_b\,\right]&=& -i\,\epsilon _{ab}{}^c S_c\,,\nonumber \\
\left[ K_a\,, K_b\,\right]&=& \,\,\,\,i\,\epsilon _{ab}{}^c S_c\,,\nonumber \\
\left[ S_a\,,K_b\,\right]&=& -i\,\epsilon _{ab}{}^c K_c\,,\nonumber \\
\left[ S_a\,,P_0\,\right]&=& \,\,\,\,0\,,\nonumber \\
\left[ S_a\,,P_b\,\right]&=& -i\,\epsilon _{ab}{}^c P_c\,,\nonumber \\
\left[ K_a\,,P_0\,\right]&=&\,\,\,\,i\,P_a\,,\nonumber \\
\left[ K_a\,,P_b\,\right]&=&\,\,\,\,i\,\delta _{ab}P_0\,,\nonumber \\
\left[ P_a\,,P_b\,\right]&=& \left[ P_a\,,P_0\,\right] =\, \left[
P_0\,,P_0\,\right] =\,0\,.\label{App2.2}
\end{eqnarray}

In the nonlinear transformation law (\ref{simplif}), we take the
infinitesimal Poincar\'e group elements $g\in G$, and the $SO(3)$
group elements $h\in H$, to be respectively
\begin{equation}
g =\,e^{i\,\epsilon^\alpha P_\alpha } e^{i\,\beta
^{\alpha\beta}L_{\alpha\beta}} \approx\,1+i\,\left(\epsilon ^0 P_0
+\epsilon ^a P_a +\zeta ^a K_a +\theta ^a
S_a\,\right)\label{App2.3}
\end{equation}
and
\begin{equation}
h =e^{i\,{\bf{\Theta}} ^a S_a}\approx\,1 +i\,{\bf{\Theta}} ^a S_a
\,,\label{App2.4}
\end{equation}
and further we parametrize $b\in G/H$ as
\begin{equation}
b=e^{-i\,\xi ^\alpha P_\alpha}e^{i\,\lambda ^a K_a}
\,,\label{App2.5}
\end{equation}
being $\xi ^\alpha \,$ and $\lambda ^a $ finite coset fields.
Eq.(\ref{simplif}) with the particular choices
(\ref{App2.3})--(\ref{App2.5}) yields on the one hand the
variation of the translational parameters
\begin{eqnarray}
\delta \xi ^0&=&-\zeta ^a \xi_a-\epsilon ^0\,,\label{App2.6a}\\
\delta \xi ^a &=&\,\,\epsilon ^a{}_{bc}\theta ^b \xi^c -\zeta ^a
\xi ^0 -\epsilon ^a\,,\label{App2.6b}
\end{eqnarray}
which, since $\zeta ^a :=\beta ^{a0}$ and $\theta ^a :=-{1\over
2}\epsilon ^a{}_{bc}\,\beta ^{bc}\,$ as read out from
(\ref{App2.3}), can be rewritten as
\begin{equation}
\delta \xi^\alpha =-\beta _\beta {}^\alpha\,\xi^\beta -\epsilon
^\alpha\,,\label{App2.7}
\end{equation}
compare with (\ref{paramvars}), showing that the coset parameters
$\xi^\alpha$ associated with the translations behave in fact as
coordinates. On the other hand, the variations of the boost
parameters in (\ref{App2.5}) turn out to be
\begin{equation}
\delta\lambda ^a =\,\epsilon ^a {}_{bc}\theta ^b \lambda ^c +\zeta
^a |\lambda |\coth |\lambda | +{{\lambda ^a\lambda _b\zeta
^b}\over{|\lambda |^2}}\left( 1-|\lambda |\coth |\lambda
|\,\right) \,,\qquad |\lambda |:=\,\sqrt{\lambda _a\lambda ^a
}\,.\label{App2.8}
\end{equation}
Instead of dealing with $\lambda ^a$, it is preferable to
introduce the velocity fields
\begin{equation}
\beta ^a :=-\,{{\lambda ^a}\over{|\lambda |}} \tanh |\lambda
|\,,\qquad \gamma :=\,{1\over{\sqrt{1-\beta ^2}}}\,,\label{App2.9}
\end{equation}
varying as
\begin{eqnarray}
\delta\,\gamma &=&-\,\zeta ^a\,\left(\gamma\,\beta _a\,\right)\,,\label{App2.11a}\\
\delta\,\left(\gamma\,\beta ^a\,\right) &=& \,\epsilon
^a{}_{bc}\,\theta ^b\left(\gamma\,\beta ^c\,\right) -\zeta
^a\,\gamma\,,\label{App2.11b}
\end{eqnarray}
that is, as the components of a Lorentz four--vector
$\left(\gamma\,,\gamma\,\beta ^a\,\right)\,$, as can be easily
checked by comparing (\ref{App2.11a}), (\ref{App2.11b}) with
(\ref{App2.6a}), (\ref{App2.6b}).

Finally, (\ref{simplif}) also enforces ${\bf{\Theta}} ^a$ in
(\ref{App2.4}) to be
\begin{equation}
{\bf{\Theta}} ^a =\,\theta ^a +{\gamma\over{\left(
1+\gamma\,\right)}} \epsilon ^a{}_{bc} \,\zeta ^b\,\beta
^c\,.\label{App2.13}
\end{equation}
According to (\ref{varmatt}), ${\bf{\Theta}} ^a$ is the modified
$SO(3)$ gauge parameter throw which the nonlinear action of the
Poincar\'e group takes place as
\begin{equation}
\delta\psi =\,i\, {\bf{\Theta}} ^a\rho\left( S_a\right)\psi
\label{App2.12}
\end{equation}
on fields $\psi$ of representation spaces of $SO(3)$, being
$\rho\left( S_a\right)$ the corresponding representation of the
$SO(3)\,$ generators.

Now we turn our attention to the nonlinear gauge fields
(\ref{compar4}), defined from the ordinary linear Poincar\'e
connection
\begin{equation}
A_{_M}:=-i\,{\buildrel (T)\over{\Gamma ^\alpha}} P_\alpha
-i\,{\buildrel {Lor}\over{\Gamma
^{\alpha\beta}}}L_{\alpha\beta}\,,\label{App2.14}
\end{equation}
standing ${\buildrel (T)\over{\Gamma ^\alpha}}$ for the
translational and ${\buildrel {Lor}\over{\Gamma ^{\alpha\beta}}}$
for the Lorentz contribution. (Although modified by this
additional specification, (\ref{App2.14}) is identical with
(\ref{linconnpoinc}).) Making use of (\ref{App2.1}) we introduce
for (\ref{compar4}) the notation
\begin{equation}
\Gamma _{_M}=-i\,\hat{\vartheta}^\alpha P_\alpha
-i\,\hat{\Gamma}^{\alpha\beta}L_{\alpha\beta}
=-i\,\hat{\vartheta}^0 P_0 -i\,\hat{\vartheta}^a P_a +i\,X^a K_a
+i\,A^a S_a \,,\label{App2.15}
\end{equation}
where a simple application of the Hausdorff-Campbell formulas of
Appendix {\bf B} yields
\begin{eqnarray}
\hat{\vartheta}^\alpha&=&\,\vartheta ^\beta b_\beta{}^\alpha\,, \label{App2.16a}\\
X^a&:=&\hat{\Gamma}^{0a}=\,(b^{-1})^{0\mu}\left( {\buildrel
{Lor}\over{\Gamma _\mu{}^\nu}}b_\nu{}^a -d\,b_\mu{}^a\,\right)\,, \label{App2.16b}\\
A^a&:=& {1\over 2}\,\epsilon ^a{}_{bc}\hat{\Gamma}^{bc}=\, {1\over
2}\,\epsilon ^a{}_{bc}\,(b^{-1})^{b\mu}\left( {\buildrel
{Lor}\over{\Gamma _\mu{}^\nu}}b_\nu{}^c
-d\,b_\mu{}^c\,\right)\,,\label{App2.16c}
\end{eqnarray}
expressed with the help of the boost matrix
\begin{eqnarray}
b_0{}^0 =(b^{-1})_0{}^0:=\gamma\,,&&\quad b_0{}^a
=-(b^{-1})_0{}^a:=-\gamma\beta ^a\,,\quad \nonumber \\
b_a{}^0 =-(b^{-1})_a{}^0:=-\gamma\beta _a\,,&&\quad b_b{}^a
=(b^{-1})_b{}^a:=\delta _b^a +(\gamma -1){{\beta _b\beta
^a}\over{\beta ^2}}\,,\hskip0.5cm\label{App2.17}
\end{eqnarray}
built from the fields (\ref{App2.9}). The Lorentz covectors
\begin{equation}
\vartheta ^\alpha :=\,{\buildrel {Lor}\over D}\xi^\alpha
+{\buildrel (T)\over{\Gamma ^\alpha}}\,,\label{App2.18}
\end{equation}
in the r.h.s. of (\ref{App2.16a}) are identical with the Lorentz
coframes (\ref{tetrad}) of the nonlinear approach studied above.
(The abbreviation $Lor$ over the covariant differentials in
(\ref{App2.18}) indicates that they are constructed with the
linear Lorentz connection ${\buildrel {Lor}\over{\Gamma
^{\alpha\beta}}}$ in (\ref{App2.14}).) Despite the formal analogy
of (\ref{App2.16a})--(\ref{App2.16c}) with gauge transformations,
in fact the coset parameters $\lambda ^a$, and thus (\ref{App2.9})
and (\ref{App2.17}), are fields of the theory rather than gauge
parameters. Consequently, (\ref{App2.16a})--(\ref{App2.16c}) are
definitions of new variables whose transformation properties
depend on (\ref{App2.11a}), (\ref{App2.11b}). Actually, while
$\vartheta ^\alpha$ in (\ref{App2.18}) transforms as a Lorentz
covector and ${\buildrel {Lor}\over {\Gamma ^{\alpha\beta}}}$ in
(\ref{App2.14}) as a Lorentz connection, for the quantities
defined in (\ref{App2.16a})--(\ref{App2.16c}) we find
\begin{eqnarray}
\delta\hat{\vartheta}^0 &=&\,0\,, \label{App2.19a}\\
\delta\hat{\vartheta}^a &=&\,\epsilon ^a{}_{bc}\,{\bf{\Theta}}
^b\, \hat{\vartheta}^c\,, \label{App2.19b}\\
\delta X^a &=&\,\epsilon ^a{}_{bc}\,{\bf{\Theta}} ^b\,X^c\,, \label{App2.19c}\\
\delta A^a &=&-D\,{\bf{\Theta}} ^a :=-\left( d\,{\bf{\Theta}} ^a
+\epsilon ^a{}_{bc}\,A^b\,{\bf{\Theta}}
^c\,\right)\,.\label{App2.19d}
\end{eqnarray}
That is, the tetrads become split into an $SO(3)$ singlet --the
invariant time component $\hat{\vartheta}^0\,$-- plus an $SO(3)$
covector --the triad $\hat{\vartheta}^a$--. The nonlinear boost
connection 1--forms $X^a$ also transform as the components of an
$SO(3)$ covector. Only the $SO(3)$ connections $A^a$ retain their
connection character. The nonlinear field strength
(\ref{fieldstrength}) built from (\ref{App2.15}) reads
\begin{equation}
F:=\,d\,\Gamma _{_M} +\Gamma _{_M}\wedge\Gamma _{_M}=-i\,\hat{T}^0
P_0 -i\,\hat{T}^a P_a +i\,(D\,X^a ) K_a+i\,{\cal R}^a S_a
\,,\label{App2.20}
\end{equation}
where we introduce the definition of the torsion
\begin{eqnarray}
\hat{T}^0 &:=&\, d\,\hat{\vartheta}^0
+\hat{\Gamma}_\mu{}^0\wedge\hat{\vartheta}^\mu
=\,d\,\hat{\vartheta}^0
+\hat{\vartheta}_a\wedge X^a\,,\label{App2.21a}\\
\hat{T}^a &:=&\,d\,\hat{\vartheta}^a
+\hat{\Gamma}_\mu{}^a\wedge\hat{\vartheta}^\mu
=\,D\,\hat{\vartheta}^a +\hat{\vartheta}^0\wedge X^a\,,\quad {\rm
with}\hskip0.2cm D\,\hat{\vartheta}^a :=\,d\,\hat{\vartheta}^a
+\epsilon ^a{}_{bc}
\,A^b\wedge\hat{\vartheta}^c\,,\label{App2.21b}
\end{eqnarray}
the boost curvature
\begin{equation}
D\,X^a :=\,d\,X^a +\epsilon ^a{}_{bc}\, A^b\wedge
X^c\,,\label{App2.22}
\end{equation}
and the rotational curvature
\begin{equation}
{\cal R}^a :=\,F^a -{1\over2}\,\epsilon ^a{}_{bc}\,X^b\wedge
X^c\,,\quad {\rm with}\hskip0.2cm F^a :=\,d\,A^a
+{1\over2}\,\epsilon ^a{}_{bc} \,A^b\wedge A^c \,,\label{App2.23}
\end{equation}
respectively. It is trivial to check
\begin{equation}
D\,X^a =\hat{R}^{0a}\,,\quad {\cal R}^a ={1\over2}\,\epsilon
^a{}_{bc}\,\hat{R}^{bc}\,,\label{App2.24}
\end{equation}
relating (\ref{App2.22}), (\ref{App2.23}) to the four--dimensional
curvature $\hat{R}_\alpha{}^\beta :=d\hat{\Gamma}_\alpha{}^\beta
+\hat{\Gamma}_\gamma{}^\beta\wedge\hat{\Gamma}_\alpha{}^\gamma\,$,
with the same form as (\ref{curvature}) but built from the Lorentz
connection $\hat{\Gamma}^{\alpha\beta}$ in (\ref{App2.15}).

\section{Foliation of several Poincar\'e objects}

In the present Appendix we apply the foliation procedure of
Subsection {\bf IV. B} to the quantities introduced in Appendix
{\bf C}. Regarding the fundamental objects
(\ref{App2.16a})--(\ref{App2.16c}), notice that trivially the zero
component of the tetrad (\ref{App2.16a}), with the form
$\hat{\vartheta} ^0 =u^0\,d\,\tau$ as in (\ref{Hamilt2}), only
includes a longitudinal contribution, whereas $\hat{\vartheta} ^a
= \underline{\hat{\vartheta}}^a$ only contains a transversal one.
On the other hand, the boost nonlinear connection (\ref{App2.16b})
and the $SO(3)$ connection (\ref{App2.16c}) decompose as $X^a
=\hat{\vartheta} ^0 X_{\bot}^a +\underline{X}^a\,$ and $A^a
=\hat{\vartheta} ^0 A_{\bot}^a +\underline{A}^a$ respectively.
Furthermore, the decomposition of the torsion components
(\ref{App2.21a}), (\ref{App2.21b}) takes the form
\begin{eqnarray}
\hat{T}^0&=&-\hat{\vartheta} ^0\wedge\left(\,\underline{d}\log u^0
+X^a_{\bot}\hat{\vartheta} _a\,\right)+\hat{\vartheta} _a\wedge
\underline{X}^a\,, \label{torsfoliat1}\\
\hat{T}^a &=&\,\hat{\vartheta} ^0\wedge\left( {\cal
\L\/}_{\hat{e}_{_0}}\hat{\vartheta} ^a +\underline{X}^a\,\right)
+\underline{D}\,\hat{\vartheta} ^a\,,\label{torsfoliat2}
\end{eqnarray}
where the covariantized Lie derivative and the transversal part of
the covariant differential are respectively defined as
\begin{equation}
{\cal \L\/}_{\hat{e}_{_0}}\hat{\vartheta}^a :=\hat{e}_{_0}\rfloor
D\hat{\vartheta} ^a ={\it{l}}_{\hat{e}_{_0}}\hat{\vartheta} ^a
+\epsilon ^a{}_{bc}A_{\bot}^b\hat{\vartheta} ^c \,,\qquad
\underline{D}\,\hat{\vartheta} ^a :=\underline{d}\,\hat{\vartheta}
^a +\epsilon ^a{}_{bc}\underline{A}^b\wedge\hat{\vartheta}
^c.\label{torsfolaux}
\end{equation}
The boost curvature (\ref{App2.22}) splits into longitudinal and
transversal parts as
\begin{equation}
D\,X^a =\hat{\vartheta} ^0\wedge\left[\,{\cal
\L\/}_{\hat{e}_{_0}}\underline{X}^a
-{1\over{u^0}}\,\underline{D}\left(
u^0\,X^a_{\bot}\,\right)\right]
+\underline{D}\,\underline{X}^a\,,\label{Hamilt26}
\end{equation}
in terms of the covariant derivatives
\begin{eqnarray}
{\cal \L\/}_{\hat{e}_{_0}}\underline{X}^a\,\,
&:=&\,\hat{e}_{_0}\rfloor D\,\underline{X}^a
=\,{\it{l}}_{\hat{e}_{_0}}\underline{X}^a
+\epsilon ^a{}_{bc}\,A_{\bot}^b\underline{X}^c\,, \label{Hamilt27a}\\
\underline{D}\left( u^0\,X^a_{\bot}\,\right)
&:=&\,\underline{d}\left( u^0\,X^a_{\bot}\,\right) +\epsilon
^a{}_{bc}\,\underline{A}^b\left( u^0\,X^c_{\bot}\right)\,, \label{Hamilt27b}\\
\underline{D}\,\underline{X}^a &:=&\underline{d}\,\underline{X}^a
+\epsilon
^a{}_{bc}\,\underline{A}^b\wedge\underline{X}^c\,,\label{Hamilt27c}
\end{eqnarray}
compare with (\ref{Hamilt7}). The $SO(3)$ curvature
(\ref{App2.23}) and its Hodge dual, see (\ref{Hamilt6}), decompose
respectively as
\begin{equation}
{\cal R}^a =\hat{\vartheta} ^0\wedge{\cal R}_{\bot}^a
+\underline{\cal R}^a\,, \qquad {}^*{\cal{R}}^a =\,\hat{\vartheta}
^0\wedge {}^\#\underline{\cal{R}}^a -{}^\#
{\cal{R}}_{\bot}^a\,,\label{Hamilt28}
\end{equation}
with definitions
\begin{equation}
{\cal{R}}_{\bot}^a :=\,F_{\bot}^a -\epsilon
^a{}_{bc}\,X_{\bot}^b\,\underline{X}^c\quad\,,\qquad F_{\bot}^a
:=\,{\it{l}}_{\hat{e}_{_0}}\,\underline{A}^a
-{1\over{u^0}}\,\underline{D}\left(\,u^0\,
A_{\bot}^a\,\right)\,,\label{Hamilt29}
\end{equation}
and
\begin{equation}
\underline{\cal R}^a :=\,\underline{F}^a -{1\over2}\,\epsilon
^a{}_{bc}\,\underline{X}^b \wedge\underline{X}^c\quad\,,\qquad
\underline{F}^a :=\,\underline{d}\,\underline{A}^a
+{1\over2}\,\epsilon ^a{}_{bc}\,\underline{A}^b\wedge
\underline{A}^c \,.\label{Hamilt30}
\end{equation}
In order to complete the set of foliated objects needed in Section
{\bf IV}, we give here the 3+1 decomposition of the 4-dimensional
eta basis of Appendix {\bf A} as
\begin{equation}
\hat{\eta}_{abc}=-\hat{\vartheta} ^0\,{\overline{\eta
}}_{abc}\,,\qquad \hat{\eta}_{ab} =\,\hat{\vartheta}
^0\wedge\overline{\eta}_{ab}\,,\qquad \hat{\eta}_{a}
=-\hat{\vartheta} ^0\wedge\overline{\eta}_a\,,\qquad \hat{\eta}
=\,\hat{\vartheta} ^0\wedge\overline{\eta}\,,\label{Hamilt32}
\end{equation}
where the bar over the etas in (\ref{Hamilt32}) means their
restriction to the three--space as
\begin{eqnarray}
{\overline{\eta }}_{abc}&:=&\hat{\eta}_{0abc}=\,\epsilon _{abc}\,,\nonumber \\
{\overline{\eta }}_{ab}&:=&\hat{\eta} _{0ab}=\,\epsilon
_{abc}\hat{\vartheta} ^c\,,\nonumber \\
{\overline{\eta }}_a &:=&\hat{\eta} _{0a}=\,{1\over{2}}\,\epsilon
_{abc} \hat{\vartheta} ^b\wedge\hat{\vartheta} ^c\,,\nonumber \\
{\overline{\eta }}&:=&\hat{\eta} _{0}=\, {1\over{3!}}\,\epsilon
_{abc}\hat{\vartheta} ^a\wedge \hat{\vartheta}
^b\wedge\hat{\vartheta} ^c\,.\label{Hamilt31}
\end{eqnarray}
The identification of ${\overline{\eta }}_{abc}$ with the group
constants $\epsilon _{abc}$ of $SO(3)$ in (\ref{Hamilt31}) is
possible due to the fact that, being the holonomic $SO(3)$ metric
the Kronecker delta, one has ${\overline{\eta }}_{abc}
={}^{\#}\left(\hat{\vartheta} _a\wedge\hat{\vartheta}
_b\wedge\hat{\vartheta} _c\, \right) =\,\sqrt{\det (\delta
_{mn})}\,\epsilon _{abc}=\, \epsilon _{abc}\,$.


\end{document}